\documentclass[twocolumn]{aastex63}
\usepackage{xcolor,soul}
\usepackage{hyperref}
\providecommand{\bjdtdb}{\ensuremath{\rm {BJD_{TDB}}}}

\providecommand{\teff}{\ensuremath{T_{\rm eff}}}

\providecommand{\msun}{\ensuremath{\,M_\Sun}}
\providecommand{\rsun}{\ensuremath{\,R_\Sun}}
\providecommand{\lsun}{\ensuremath{\,L_\Sun}}
\providecommand{\mj}{\ensuremath{\,{\rm M_J}}}
\providecommand{\rj}{\ensuremath{\,{\rm R_J}}}

\providecommand{\mst}{\ensuremath{\,{\rm M_\odot}}}
\providecommand{\rst}{\ensuremath{\,{\rm R_\odot}}}

\providecommand{\arcsec}{$^{\prime \prime}$}
\providecommand{\toibmass}{$64.1 \pm 1.9$}
\providecommand{\toibrad}{$0.75 \pm 0.02$}
\providecommand{\toimass}{$1.21 \pm 0.05$}
\providecommand{\toirad}{$1.47 \pm 0.03$}
\providecommand{\toitemp}{$5768 \pm 110$}

\providecommand{\ticbmass}{$46.0 \pm 2.7$}
\providecommand{\ticbrad}{$0.86 \pm 0.03$}
\providecommand{\ticmass}{$1.18 \pm 0.09$}
\providecommand{\ticrad}{$1.35 \pm 0.03$}
\providecommand{\tictemp}{$6290 \pm 100$}

\providecommand{\dbf}{}
\newcommand\ajm[1]{\textcolor{blue}{\textbf{#1}}}
\newcommand\dsout[1]{}

\graphicspath{{./}{figures/}}

\begin{document}

\title{Two intermediate-mass transiting brown dwarfs from the TESS mission}

\author[0000-0001-6416-1274]{Theron W. Carmichael}
\affil{\rm Department of Astronomy, Harvard University,
Cambridge, MA 02138 \\ email: \rm \href{mailto:tcarmich@cfa.harvard.edu}{tcarmich@cfa.harvard.edu}}
\affil{\rm Center for Astrophysics ${\rm \mid}$ Harvard {\rm \&} Smithsonian, 60 Garden Street, Cambridge, MA 02138, USA}
\affil{\rm National Science Foundation Graduate Research Fellow}

\author{Samuel N. Quinn}
\affil{\rm Center for Astrophysics ${\rm \mid}$ Harvard {\rm \&} Smithsonian, 60 Garden Street, Cambridge, MA 02138, USA}

\author{Alexander J. Mustill}
\affil{\rm Lund Observatory, Department of Astronomy and Theoretical Physics, Lund University, Box 43, SE-221 00 Lund, Sweden}

\author{Chelsea Huang}
\affil{\rm Department of Physics, and Kavli Institute for Astrophysics and Space Research, Massachusetts Institute of Technology, Cambridge, MA 02139, USA}

\author{George Zhou}
\affil{\rm Center for Astrophysics ${\rm \mid}$ Harvard {\rm \&} Smithsonian, 60 Garden Street, Cambridge, MA 02138, USA}

\author{Carina M. Persson}
\affil{\rm Chalmers University of Technology, Department of Space, Earth and Environment, Onsala Space Observatory, SE-439 92 Onsala, Sweden}

\author{Louise D. Nielsen}
\affil{\rm Geneva Observatory, University of Geneva, Chemin des Mailettes 51, 1290 Versoix, Switzerland}

\author[0000-0001-6588-9574]{Karen A.\ Collins} 
\affiliation{\rm Center for Astrophysics ${\rm \mid}$ Harvard {\rm \&} Smithsonian, 60 Garden Street, Cambridge, MA 02138, USA}

\author{Carl Ziegler}
\affil{\rm Dunlap Institute for Astronomy and Astrophysics, University of Toronto, 50 St. George Street, Toronto, Ontario M5S 3H4, Canada}

\author[0000-0003-2781-3207]{Kevin I.\ Collins}
\affiliation{\rm George Mason University, 4400 University Drive, Fairfax, VA, 22030 USA}

\author{Joseph E. Rodriguez}
\affil{\rm Center for Astrophysics ${\rm \mid}$ Harvard {\rm \&} Smithsonian, 60 Garden Street, Cambridge, MA 02138, USA}

\author{Avi Shporer}
\affil{\rm Department of Physics, and Kavli Institute for Astrophysics and Space Research, Massachusetts Institute of Technology, Cambridge, MA 02139, USA}

\author{Rafael Brahm}
\affil{\rm Facultad de Ingenier\'ia y Ciencias, Universidad Adolfo Ib\'a\~{n}ez, Av.Diagonal las Torres 2640, Pe\~{n}alol\'en, Santiago, Chile}
\affil{\rm Millennium Institute for Astrophysics, Chile}

\author[0000-0003-3654-1602]{Andrew W. Mann}
\affiliation{\rm Department of Physics and Astronomy, The University of North Carolina at Chapel Hill, Chapel Hill, NC 27599-3255, USA}

\author{Francois Bouchy}
\affil{\rm Geneva Observatory, University of Geneva, Chemin des Mailettes 51, 1290 Versoix, Switzerland}

\author{Malcolm Fridlund}
\affil{\rm Chalmers University of Technology, Department of Space, Earth and Environment, Onsala Space Observatory, SE-439 92 Onsala, Sweden}
\affil{\rm Leiden Observatory, University of Leiden, PO Box 9513, 2300 RA, Leiden, The Netherlands}

\author[0000-0002-3481-9052]{Keivan G.\ Stassun}
\affil{\rm Vanderbilt University, Department of Physics \& Astronomy, 6301 Stevenson Center Ln., Nashville, TN 37235, USA}
\affil{\rm Fisk University, Department of Physics, 1000 18th Ave. N., Nashville, TN 37208, USA}

\author{Coel Hellier}
\affil{\rm Astrophysics Group, Keele University, Staffordshire, ST5 5BG, UK}

\author{Julia V. Seidel}
\affil{\rm Geneva Observatory, University of Geneva, Chemin des Mailettes 51, 1290 Versoix, Switzerland}

\author{Manu Stalport}
\affil{\rm Geneva Observatory, University of Geneva, Chemin des Mailettes 51, 1290 Versoix, Switzerland}

\author{Stephane Udry}
\affil{\rm Geneva Observatory, University of Geneva, Chemin des Mailettes 51, 1290 Versoix, Switzerland}

\author{Francesco Pepe}
\affil{\rm Geneva Observatory, University of Geneva, Chemin des Mailettes 51, 1290 Versoix, Switzerland}

\author{Michael Ireland}
\affil{\rm Research School of Astronomy and Astrophysics, The Australian National University, ACT, 2611}

\author{Maru{\v s}a {\v Z}erjal}
\affil{\rm Research School of Astronomy and Astrophysics, The Australian National University, ACT, 2611}

\author{C\'{e}sar Brice\~{n}o}
\affiliation{\rm Cerro Tololo Inter-American Observatory, Casilla 603, La Serena, Chile} 

\author{Nicholas Law}
\affiliation{\rm Department of Physics and Astronomy, The University of North Carolina at Chapel Hill, Chapel Hill, NC 27599-3255, USA}

\author{Andr\'es Jord\'an}
\affil{\rm Facultad de Ingenier\'ia y Ciencias, Universidad Adolfo Ib\'a\~{n}ez, Av.Diagonal las Torres 2640, Pe\~{n}alol\'en, Santiago, Chile}
\affil{\rm Millennium Institute for Astrophysics, Chile}

\author{N\'estor Espinoza}
\affil{\rm Space Telescope Science Institute, 3700 San Martin Drive, Baltimore, MD 21218, USA}

\author{Thomas Henning}
\affil{\rm Max-Planck-Institut f\"ur Astronomie, K\"onigstuhl 17, Heidelberg 69117, Germany}

\author{Paula Sarkis}
\affil{\rm Max-Planck-Institut f\"ur Astronomie, K\"onigstuhl 17, Heidelberg 69117, Germany}

\author{David W. Latham}
\affil{\rm Center for Astrophysics ${\rm \mid}$ Harvard {\rm \&} Smithsonian, 60 Garden Street, Cambridge, MA 02138, USA}

\begin{abstract}
\noindent We report the discovery of two intermediate-mass \dbf{transiting} brown dwarfs (BDs), TOI-569b and TOI-1406b, from NASA's Transiting Exoplanet Survey Satellite mission. TOI-569b has an orbital period of $P = 6.55604 \pm 0.00016$ days, a mass of $M_b =$ \toibmass$\mj$, and a radius of $R_b = $ \toibrad$\rj$. Its host star, TOI-569, has a mass of $M_\star = $ \toimass$\mst$, a radius of $R_\star =$ \toirad$\rst$, \dbf{$\rm   [Fe/H] = +0.29 \pm 0.09$ dex,} and an effective temperature of $T_{\rm eff} =$ \toitemp K. TOI-1406b has an orbital period of $P = 10.57415 \pm 0.00063$ days, a mass of $M_b =$ \ticbmass$\mj$, and a radius of $R_b = $ \ticbrad$\rj$. The host star for this BD has a mass of $M_\star =$ \ticmass$\mst$, a radius of $R_\star = $ \ticrad$\rst$, \dbf{$\rm   [Fe/H] = -0.08 \pm 0.09$ dex,} and an effective temperature of $T_{\rm eff} = $ \tictemp K. Both BDs are in circular orbits around their host stars and \dbf{are older than 3 Gyr based on stellar isochrone models of the stars.} \dsout{join an increasing number of known transiting intermediate-mass BDs.} TOI-569 is one of two slightly evolved stars known to host a transiting BD (the other being KOI-415). TOI-1406b is one of three known transiting BDs to occupy the mass range of 40-50$\mj$ and one of two to have a circular orbit at a period near 10 days (with the first being KOI-205b). \dsout{Based on the relatively long circularization timescales for both BDs, we believe that they must have formed in nearly circular orbits and migrated inward to their present orbital configurations.} \dbf{Both BDs have reliable ages from stellar isochrones in addition to their well-constrained masses and radii, making them particularly valuable as tests for substellar isochrones in the BD mass-radius diagram.} 
\end{abstract}


\section{Introduction} \label{sec:intro}
Brown dwarfs (BDs) are traditionally defined as objects with masses between 13 and 80 Jupiter masses ($\mj$) and, for those that orbit main-sequence stars, are typically observed to be 0.7 to 1.4 Jupiter radii ($\rj$) in size \citep{csizmadia16, carmichael19}. The lower mass limit that separates planets from BDs corresponds to the threshold required to fuse deuterium in the core of the BD, \dbf{which, in detail, is $  11-16\mj$ depending on the metallicity of the BD \citep{spiegel2011}}. \dbf{The threshold to fuse hydrogen in the core is $  75-80\mj$ and this separates BDs from stars. This higher threshold depends on the initial formation conditions and how convection affects the object \citep{baraffe2002}.} The mass and radius of a BD are measured through a combination of observational techniques, with the most important being radial velocity (RV) measurements and transit photometry of the host star. 

\dbf{This is where the Transiting Exoplanet Survey Satellite (TESS) mission plays a major role.} The transit method has been most successful in the characterization of BDs in relatively short orbital periods (\dbf{on the order of 10 to 20 days or less} to detect multiple transits), which is why the TESS mission has been particularly useful in making the initial detections of recently discovered transiting BDs \citep[e.g.][]{subjak2019, jackman2019}. The transit light curves from TESS are taken over roughly 28 consecutive days per sector (or up to one year for overlapping sectors), with occasional gaps in coverage due to data downloads and instrumental anomalies. These light curves give an estimate of the radius of the candidate companions \dbf{relative to the radius of the star}. This informs us on whether or not a candidate companion is within the typical range of radii expected for a BD orbiting a main-sequence star. We are particularly fortunate at the present time to be able to utilize the parallax measurements from \textit{Gaia} DR2 \dbf{\citep{gaiaDR2}} to derive precise stellar distances and radii for stars that host transiting BDs.

Though TESS and \textit{Gaia} provide us with some handle on the radius of a BD, follow up spectra are needed to accumulate a series of RV measurements to determine an orbit and measure a mass. This is an important step as objects ranging from roughly $1\mj$ to $100\mj$ may have the same radius ($\sim$1$\rj$), so the only way to distinguish them is through a mass measurement. \dbf{Since the masses of transiting BDs produce large RV signals around typical FGK main sequence stars, RV follow up of these BDs is fairly accessible (especially given the precision of modern echelle spectrographs).}

\dbf{When comparing the detection of transiting BDs to the detection of transiting giant planets, we see two facts emerge: 1) for host stars with similar radii, transiting BDs should be roughly as easy to detect as hot Jupiters given both types of objects are similar in size and given the photometric precision and sensitivity of transit survey missions like TESS. 2) for host stars of similar masses, it is as easy or easier to characterize the mass of a BD given that they are more massive and produce larger RV signals than giant planets.} Despite this, we know of only 23 transiting BDs (see Table \ref{tab:bdlist} for a list). When compared to the number of known transiting hot Jupiters and even eclipsing low-mass stellar companions in comparable orbital periods, it is easy to see that there is a lack of \dbf{transiting} BDs; this is the so-called ``brown dwarf desert" \citep{bdd2000}. \dbf{Though not yet clear, this ``desert" may result from the distributions of the two different formation mechanisms (one for planets and one for stars) tailing off---and perhaps overlapping---somewhere within the nominal BD mass range.}

\dsout{Despite not fully understanding the origins of this desert, we may still strive to understand the nature of the population of the objects \dsout{in} \dbf{spanning} the $13\mj$ to $80\mj$ mass range. Ma \& Ge 2014 contribute to this by proposing the existence of a gap within the the brown dwarf desert centered on $42.5\mj$, implying that two sub-populations of BDs exist above and below this mass. Another study by Persson et al. 2019 implies that the mass range of giant planets spans 0.3-73$\mj$ based on a mass-density relation. These differences in ideas highlight the importance of the detection and characterization of new transiting BDs. With a transit, orbital solution, and well-determined stellar radius, we have some of the key components required to precisely measure the mass and radius of a transiting BD. So, each newly discovered transiting BD will add another member to this scarcely populated region of objects.}

\dbf{To understand this population on a deeper level, we use} the mass, radius, \dbf{and age} of transiting BDs to examine how these substellar objects evolve compared to substellar evolutionary models \citep[e.g.][]{baraffe03}. Since a BD may only fuse deuterium and not hydrogen, it lacks the energy source needed to stave off gravitational contraction on long timescales as effectively as stars do, so the BD's radius will shrink as it ages. \dbf{This is why the age is important in addition to mass and radius. If we can reliably determine the age of a star that hosts a transiting BD, whether through an association with a star cluster, gyrochronology, or asteroseismology, then we may use that transiting BD to test substellar evolutionary models in mass, radius, and age parameter spaces. This assumes that both the host star and the transiting BD form at the same time. So far, only 3 BDs transiting main sequence stars with well-determined ages are known \citep{cww89a, david19_bd, ad3116}, so we are lacking test points on the substellar mass-radius diagram with precisely determined mass, radius, and age.}

\begin{deluxetable}{lcccc}
 \tabletypesize{\footnotesize}
 \tablewidth{0pt}
 \tablecaption{Coordinates and magnitudes for TOI-569 and TOI-1406. \dbf{The values here make up the spectral energy distributions used to model the stars.} \label{tab:toi_obs}}
 \tablehead{
 \colhead{} & \colhead{Description} & \colhead{TOI-569} & \colhead{TOI-1406} & \colhead{Source}}
 \startdata
 $\alpha_{\rm J2000}$ &Equatorial& 07 40 24.67 & 05 28 30.71 & 1\\
 $\delta_{\rm J2000}$ &coordinates& -42 09 16.79 & -48 24 32.64 & 1\\
 \smallskip\\
 $T$\dotfill & TESS $T$\dotfill & $9.473 \pm 0.006$ & $11.427 \pm 0.006$ & 2\\
 $G$\dotfill & Gaia $G$\dotfill & $9.936 \pm 0.001$ & $11.759 \pm 0.001$ & 1\\
 $B_T$\dotfill & Tycho $B_T$\dotfill & $11.036 \pm 0.048$ & $13.352 \pm 0.359$ & 3\\
 $V_T$\dotfill & Tycho $V_T$\dotfill & $10.173 \pm 0.032$ & $12.074 \pm 0.195$ & 3\\
 $J$\dotfill & 2MASS $J$\dotfill & $8.829 \pm 0.020$ & $10.929 \pm 0.020$ & 4\\
 $H$\dotfill & 2MASS $H$\dotfill & $8.575 \pm 0.060$ & $10.787 \pm 0.030$ & 4\\
 $K_S$\dotfill & 2MASS $K_S$\dotfill & $8.444 \pm 0.020$ & $10.675 \pm 0.020$ & 4\\
 WISE1\dotfill & WISE 3.4$\rm \mu m$\dotfill & $8.419 \pm 0.023$ & $10.649 \pm 0.023$  & 5\\
 WISE2\dotfill & WISE 4.6$\rm \mu m$\dotfill & $8.467 \pm 0.020$ & $10.692 \pm 0.021$ & 5\\
 WISE3\dotfill & WISE 12$\rm \mu m$\dotfill & $8.414 \pm 0.021$ & $10.651 \pm 0.062$ & 5\\
 WISE4\dotfill & WISE 22$\rm \mu m$\dotfill & $8.180 \pm 0.188$ & - & 5\\
 \hline
 \enddata
 \tablecomments{References: 1 - \cite{Lindegren2018}, corrected from the J2015 epoch, 2 - \cite{stassun18}, 3 - \cite{tycho2}, 4 - \cite{2MASS}, 5 - \cite{WISE}}
\end{deluxetable}

\begin{figure*}[!ht]
\centering
\includegraphics[width=0.99\textwidth, trim={0.0cm 0.0cm 0.0cm 0.0cm}]{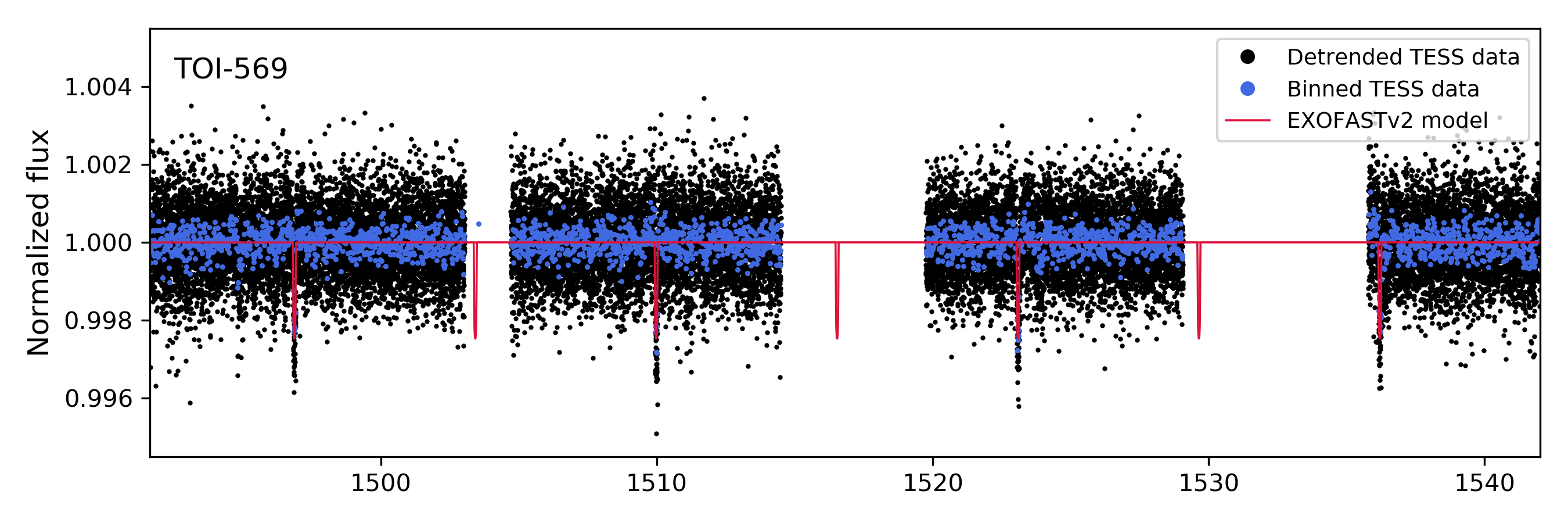}
\includegraphics[width=0.99\textwidth, trim={0.0cm 0.0cm 0.4cm 0.0cm}]{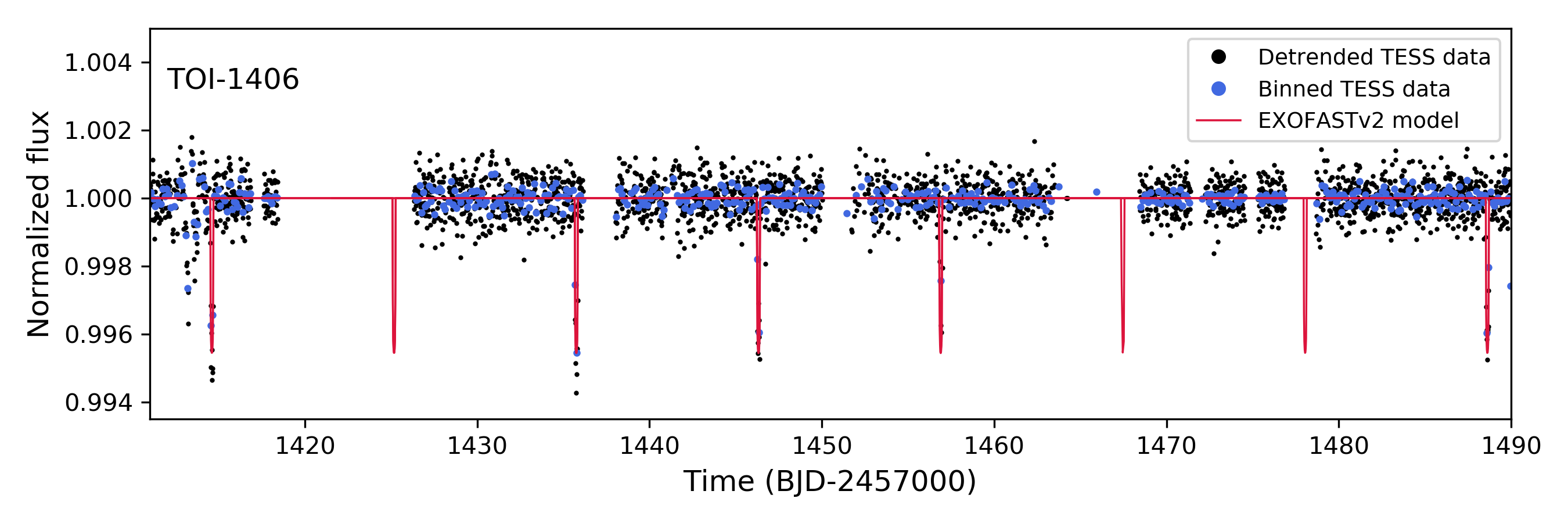}
\caption{\textit{Top}: Detrended TESS light curve of TOI-569 in the black points. The star was observed at 2 minutes cadence in TESS sectors 7 and 8 \dbf{and the binning shown here in the blue points uses bin sizes of 45 minutes.} This star also exhibits periodic 1-3\% flux variations likely due to star spots \dbf{based on the changes in the patterns of the modulation}; these effects have been removed for the transit analysis. \textit{Bottom}: Detrended TESS light curve of TOI-1406 (black points) obtained from the full-frame images at 30 minute cadence from TESS sectors 4, 5, and 6 \dbf{with the bin sizes at a length of 90 minutes (blue points).}}\label{fig:lc_detrend}
\end{figure*}

\dsout{Age is a particularly difficult parameter to measure for stars and BDs. This makes it all the more important that we have some reasonable estimate of the age as the age of a transiting BD may be compared to its circularization timescale for the system it is in. These timescales, though not well-understand for short-period BDs, are used to understand a BD’s formation and past orbital evolution.}

\dsout{One method to measure the age of a transiting BD is done via the BD's association with a star cluster. There are only 3 transiting BD systems that are in clusters or stellar associations, so in general, this population \dsout{is lacking} \dbf{lacks} systems with directly measured ages. If we are not fortunate enough to have a directly measurable age, then we rely on stellar isochrone models in combination with parallax measurements from the \textit{Gaia} mission and measurements of the spectral energy distribution (SED) of the host star. This method is limited by how well the metallicity and effective temperature of the star is measured.}

Here we report the discovery and characterization of two new transiting BDs \dbf{with reliable mass, radius, and age measurements: TOI-569b and TOI-1406b.} TOI-569b orbits a recently evolved star in a circular orbit. TOI-1406b orbits an F-type star, joining 7 other transiting BDs that orbit F-type stars, and is also in a circular orbit. \dbf{These both serve as new test points on the mass-radius diagram for BDs older than 3 Gyr.} In Section \ref{sec:observations}, we give details on the light curves and spectra that were obtained for this study with additional attention given to the determination of the orbital period of TOI-569b, which was initially reported incorrectly \dbf{as twice its actual value} due to gaps in the TESS data. Section \ref{sec:analysis} describes the analysis techniques used to derive the host star and BD properties. Section \ref{sec:conclusion} contains discussion of the implications of these new discoveries in the BD mass-radius diagram \dbf{as these two transiting BDs are the oldest transiting BDs that have been well-characterized.} \dsout{and some discussion on the circularization timescales for each system.} 

\section{Observations}\label{sec:observations}
\subsection{TESS and ground-based light curves}
The light curves of TOI-569 come from the TESS mission in sectors 7 and 8, 
and the Las Cumbres Observatory (LCOGT). For the TESS light curve for TOI-569, we use the Pre-search Data Conditioning Simple Aperture Photometry flux \citep[PDCSAP;][]{stumpe2014_pdc, smith2012_pdc} from the Mikulski Archive for Space Telescopes (MAST)\footnote{\url{https:// mast.stsci.edu/portal/Mashup/Clients/Mast/Portal.html}}. The PDCSAP light curve has systematic effects removed and with this, we then normalize \dbf{the} light curve with the {\tt lightkurve} package in Python \citep[the][]{lightkurve}. The light curves used for TOI-1406 come from the full-frame images (30 minute cadence) from the TESS mission in sectors 4, 5, and 6. We use the {\tt lightkurve} package to extract and normalize the light curve of TOI-1406. 

\dbf{For the light curve extraction, we use circular apertures that are fixed to the target star's position for each sector. When the star moves slightly between sectors, the aperture is moved to follow it. The counts from each pixel within the aperture are summed and the resulting light curve is detrended using the {\tt lightkurve} package's built-in flattening tool, which we use to remove stellar rotational variability, when present, as well as scattered background light. The detrended TESS light curves are shown in Figure \ref{fig:lc_detrend}.}

We observed an ingress of TOI-569 continuously for 140 minutes on April 15, 2019 using 15~s exposures and a z-short band filter from the LCOGT \citep{Brown:2013} 1.0\,m node at Cerro Tololo Inter-American Observatory. We used the {\tt TESS Transit Finder}, which is a customized version of the {\tt Tapir} software package \citep{Jensen:2013}, to schedule our transit observations. The $4096\times4096$ LCOGT SINISTRO cameras have an image scale of 0.389$\arcsec$ per pixel, resulting in a $26\arcmin\times26\arcmin$ field of view. The images were calibrated by the standard LCOGT {\tt BANZAI} pipeline, and photometric data were extracted with the {\tt AstroImageJ} software package \citep{Collins:2017} using a circular aperture with radius 5.8$\arcsec$. The images have typical stellar point-spread-functions with a half-width-half-maximum of 1$\arcsec$. We detect a $\sim3000$ ppm ingress on target with apertures as small as 2$\arcsec$ in radius. Systematic effects start to dominate the light curve for smaller apertures. Thus, we confirm that the source of the TESS detection is within 3$\arcsec$ of the target star location and that the transit depth from the LCOGT partial transit is consistent with the TESS depth for all aperture radii we checked down to 2$\arcsec$. We did not obtain any ground-based photometric followup of TOI-1406.

\begin{figure}[!ht]
\centering
\includegraphics[width=0.45\textwidth, trim={0.5cm 0.0cm 0.0cm 0.0cm}]{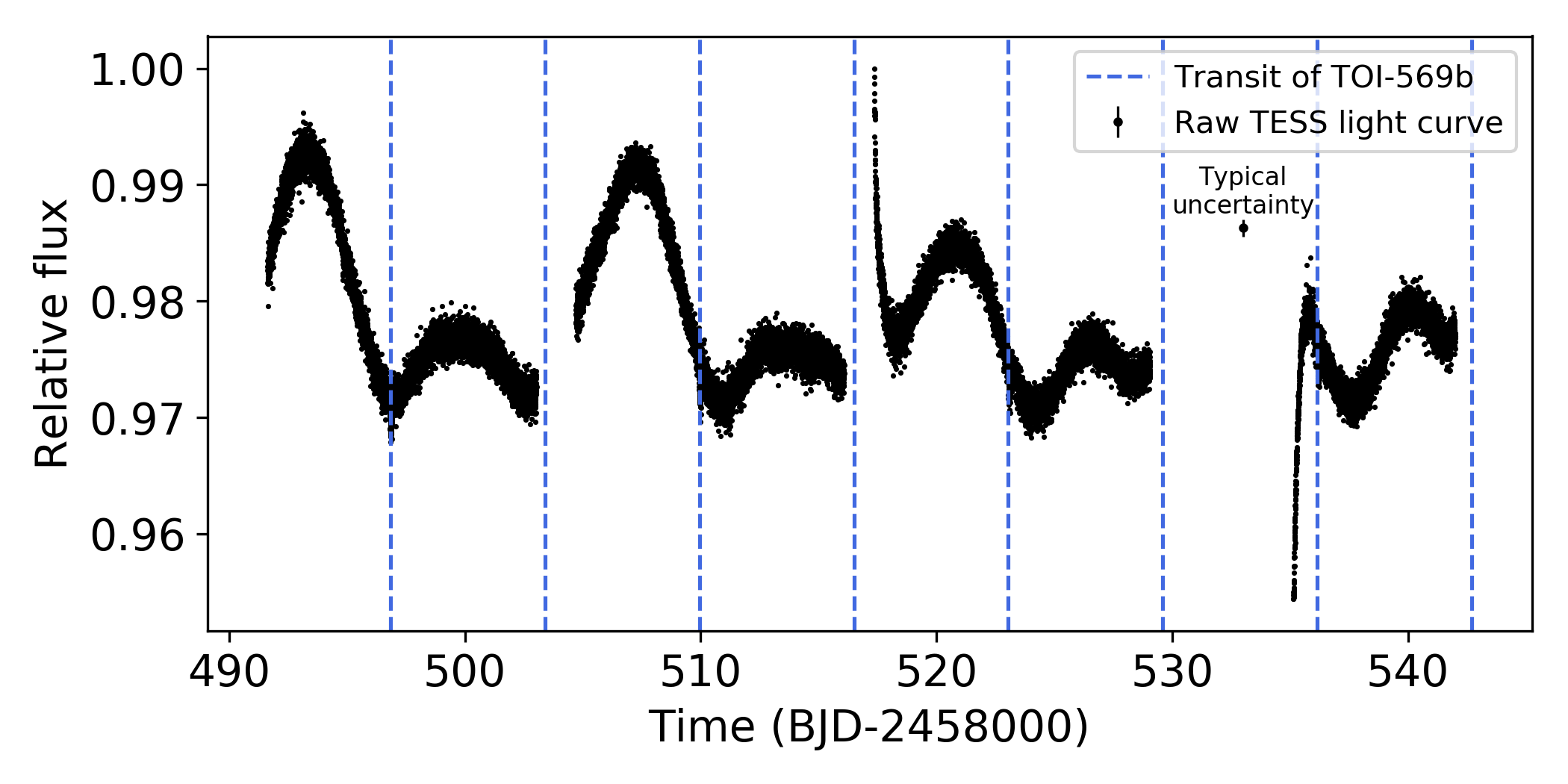}
\includegraphics[width=0.45\textwidth, trim={1.0cm 0.0cm 0.0cm 0.0cm}]{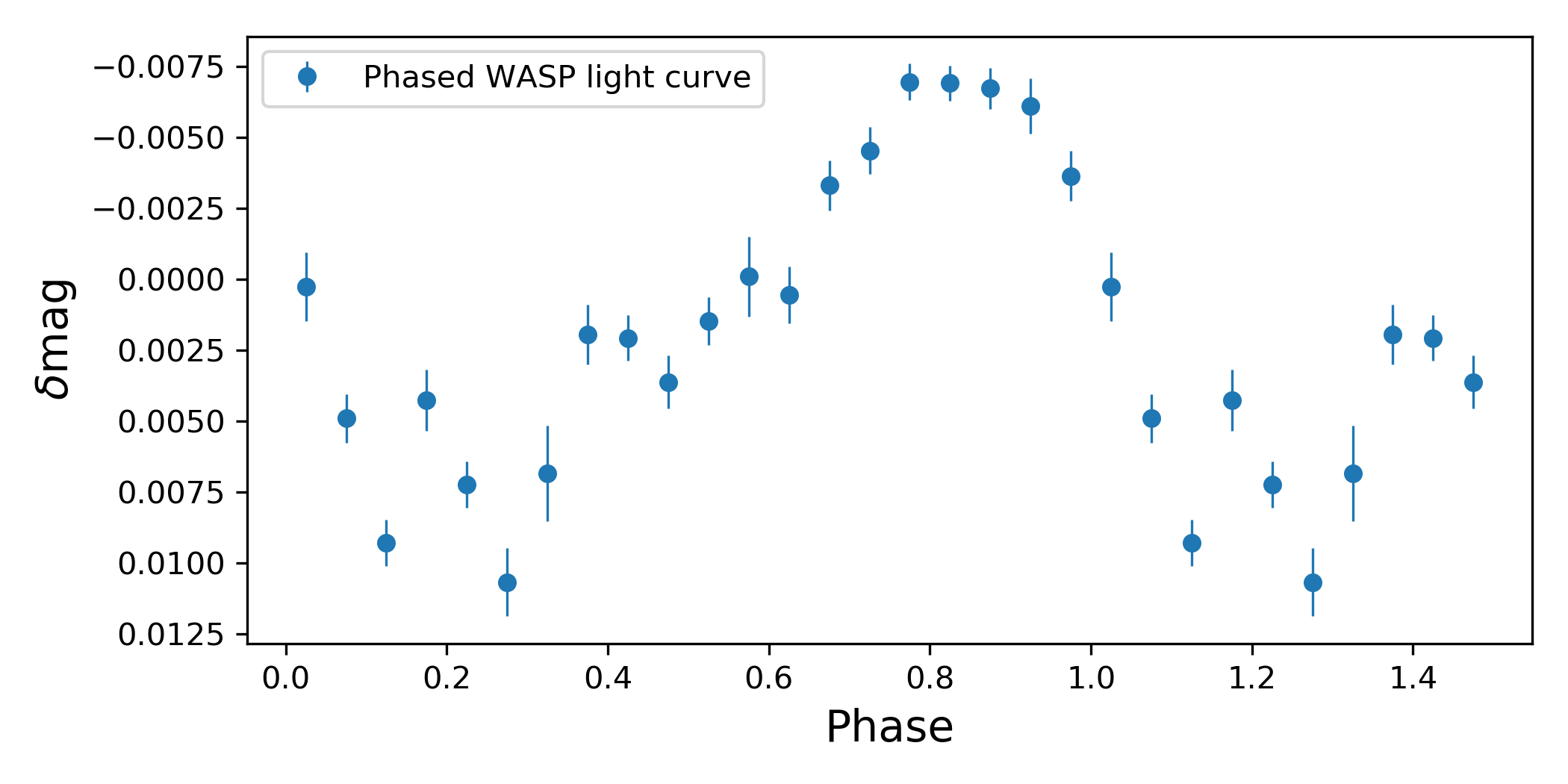}
\caption{\textit{Top}: Offset-normalized raw TESS light curve from sectors 7 and 8 of TOI-569. The offset between each sector is removed using {\tt lightkurve}, but no other systematic effects are removed since the focus of this figure is to show the missed transits of TOI-569b and provide context for the flux variability of the host star. The rapid ramp down at 515 days and ramp up at 535 days are instrumental systematics from the spacecraft. The blue dashed lines show the transit times predicted by the final model ephemeris for TOI-569b. \textit{Bottom}: WASP light curve of TOI-569 phase folded at 13.03 days from observations taken over a span of 150 days.}\label{fig:wasp}
\end{figure}

\begin{figure}[!ht]
\centering
\includegraphics[width=0.45\textwidth, trim={0.5cm 0.0cm 0.0cm 0.0cm}]{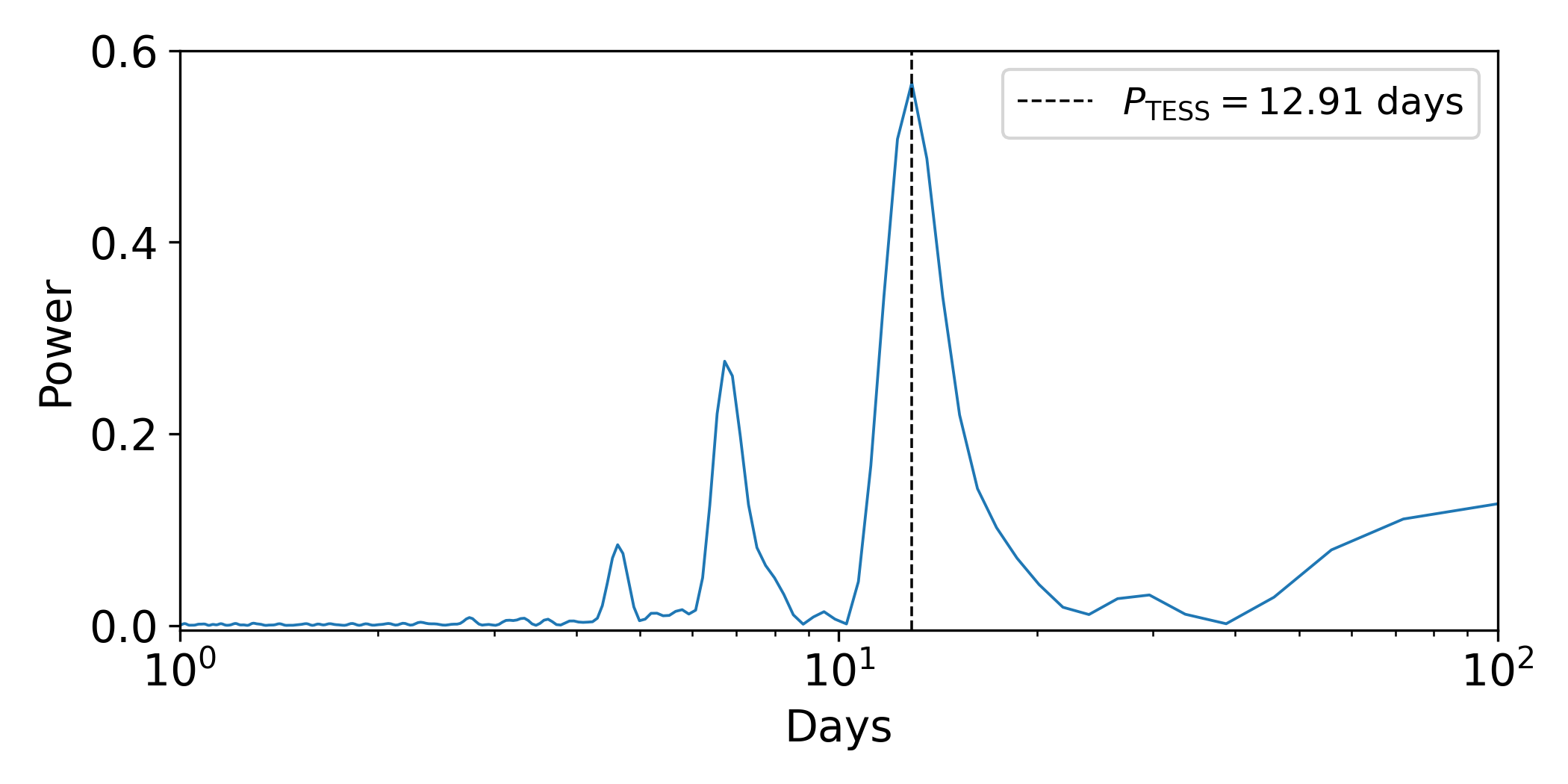}
\includegraphics[width=0.45\textwidth, trim={1.0cm 0.0cm 0.0cm 0.0cm}]{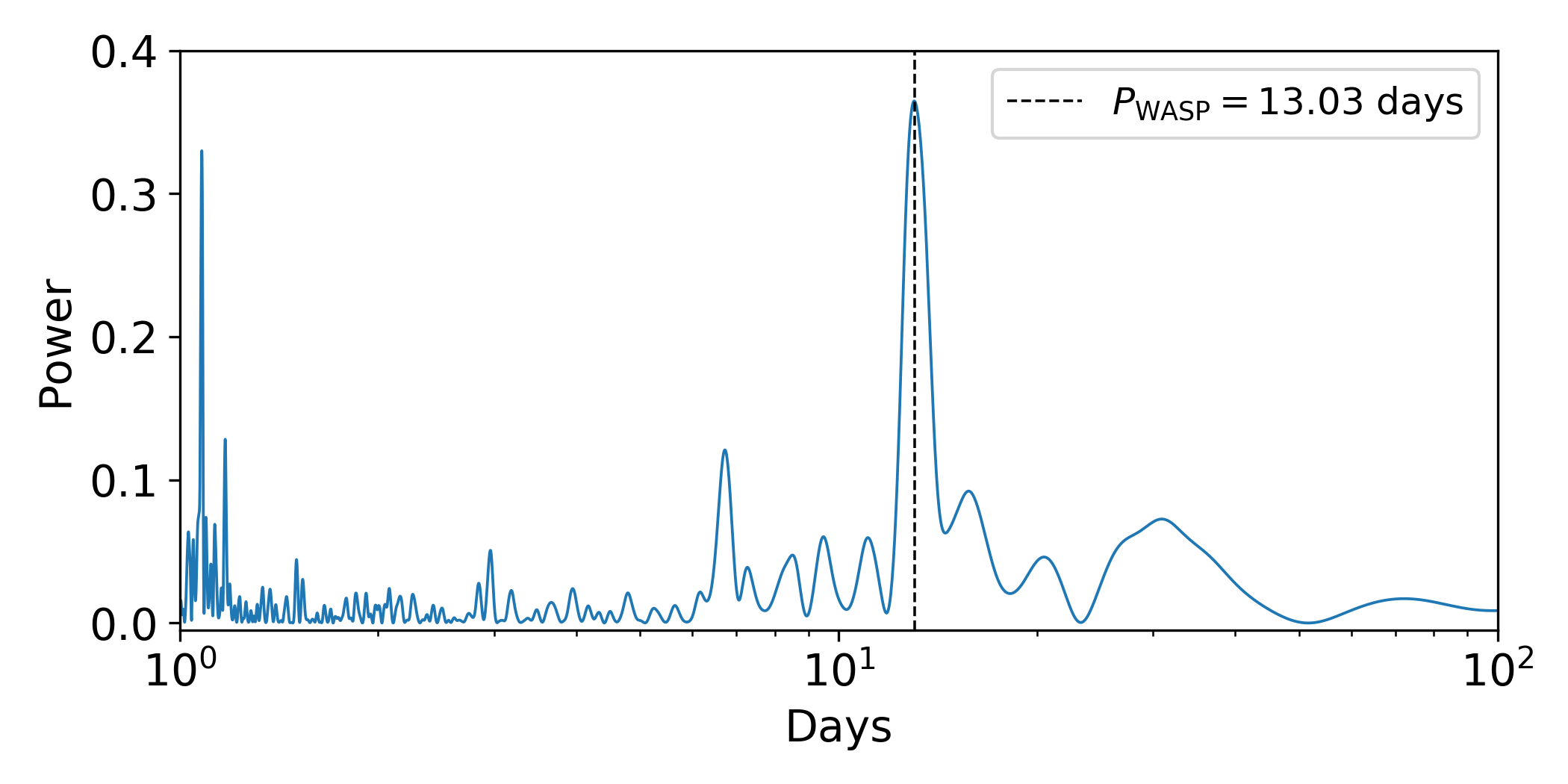}
\caption{\textit{Top}: Lomb-Scargle periodogram from the TESS light curve of TOI-569. \textit{Bottom}: Lomb-Scargle periodogram from the WASP light curve of TOI-569. Both the WASP and TESS periodograms indicate a peak frequency near 13 days. The peak in the lower panel is narrower because the total time coverage from the WASP data was nearly 3 times longer than that from TESS.}\label{fig:periodogram}
\end{figure}

\subsubsection{Light curve modulation and the orbital period of the TOI-569 system}
Previous to the transit detections of TOI-569b from TESS, the Wide Angle Search for Planets (WASP) found a 13-day modulation in the light curve of TOI-569. The phased light curve for WASP is shown in Figure~\ref{fig:wasp}. The WASP data were taken during the 2011 to 2012 seasons with a 150-day span of coverage. The transits of TOI-569b are too shallow to be detected in the WASP light curve \dbf{even when phase folded to the ephemeris from TESS}, but they can be seen in the TESS light curve in the top panel of Figure~\ref{fig:wasp}. The WASP and TESS light curves show a similar but not exactly equal modulation. The periodic peaks and dips in both light curves are likely from brightness variations due to star spots, which are known to vary over time. \dbf{The unevenness in the peaks in the TESS light curves are likely from multiple different star spot configurations on different areas on the surface of the star in addition to the evolution of spot brightness over time.}

The gaps in the TESS light curve occur during every other transit of TOI-569b in the sectors the host star was observed. This is the reason why the initial orbital period was reported to be 13.12 days (twice the true orbital period of 6.56 days). We discovered that this was the case as the orbital solution developed with RV follow up using the instruments described in later sections. It seems coincidental that this erroneous orbital period of 13.12 days is nearly equal to the 13-day modulation in the WASP light curve. This made the BD appear to have an orbit synchronized with the rotation rate of the star, but this turned out not to be the case upon a more thorough investigation that accounts for the orbital solution derived from RVs. \dbf{We note that the observation of the transit of TOI-569b with the LCOGT 1-meter telescope did not occur at opposite parity to the transits detected by TESS, so we cannot use this to independently confirm the 6.56-day period of TOI-569b.}

\dbf{Using a Lomb-Scargle periodogram analysis on both the TESS and WASP light curves separately, we see a peak frequency at roughly 13 days (12.91 days for TESS and 13.01 days for WASP). This may suggest that TOI-569 and its companion BD are in a 2:1 spin-orbit resonance.}

\begin{deluxetable*}{ccccccc}
\tabletypesize{\footnotesize}
\tablewidth{0pt}

 \tablecaption{Nearby sources from \textit{Gaia} DR2 data. This table list sources within $30\arcsec$ of each star (TOI-569 and TOI-1406) that are $G<16$ in magnitude. Listing sources fainter than this results in too many items to reasonably list here. The parallaxes ($\pi$) and proper motions ($\mu_{\rm \alpha}$, $\mu_{\rm \delta}$) of the nearby stars indicate that none are associated with TOI-569 or TOI-1406. \label{tab:gaia}}

 \tablehead{
 \colhead{\textit{Gaia} DR2 ID} & \colhead{$\alpha$ (J2000)} & \colhead{$\delta$ (J2000)}  & \colhead{$\pi$ (mas)} & \colhead{$\mu_{\rm \alpha}$ (mas/yr)} & \colhead{$\mu_{\rm \delta}$ (mas/yr)} & $G$ (mag)}

\startdata 
5535473358555685760 (TOI-569) & 07 40 24.67 & -42 09 16.79 & $6.3723 \pm 0.0306$ & $6.317 \pm 0.053$ & $-3.068 \pm 0.048$ & 9.94 \\
5535473392915426304 & 07 40 26.44 & -42 09 00.02 & $0.0899 \pm 0.0193$ & $-1.790 \pm 0.035$ & $2.832 \pm 0.030$ & 13.69 \\
5535473358555686656 & 07 40 23.61 & -42 09 39.57 & $0.8838 \pm 0.0275$ & $12.130 \pm 0.047$ & $-13.344 \pm 0.041$ & 14.88 \\
5535473358556195840 & 07 40 23.25 & -42 09 38.38 & $0.1963 \pm 0.0414$ & $-4.600 \pm 0.072$ & $5.585 \pm 0.060$ & 15.64 \\
\hline
4797030079342886784 (TOI-1406) & 05 28 30.71 & -48 24 32.64 & $2.3855 \pm 0.0291$ & $0.889 \pm 0.057$ & $-21.885 \pm 0.066$ & 11.76 \\
4797030079342886656 & 05 28 29.07 & -48 24 41.93 & $1.1846 \pm 0.0396$ & $1.222 \pm 0.073$ & $-1.752 \pm 0.093$ & 15.78 \\
\enddata
\end{deluxetable*}

\subsection{High resolution imaging and contaminating sources}
\begin{figure}[!ht]
\centering
\includegraphics[width=0.45\textwidth, trim={0.0cm 0.0cm 0.0cm 0.0cm}]{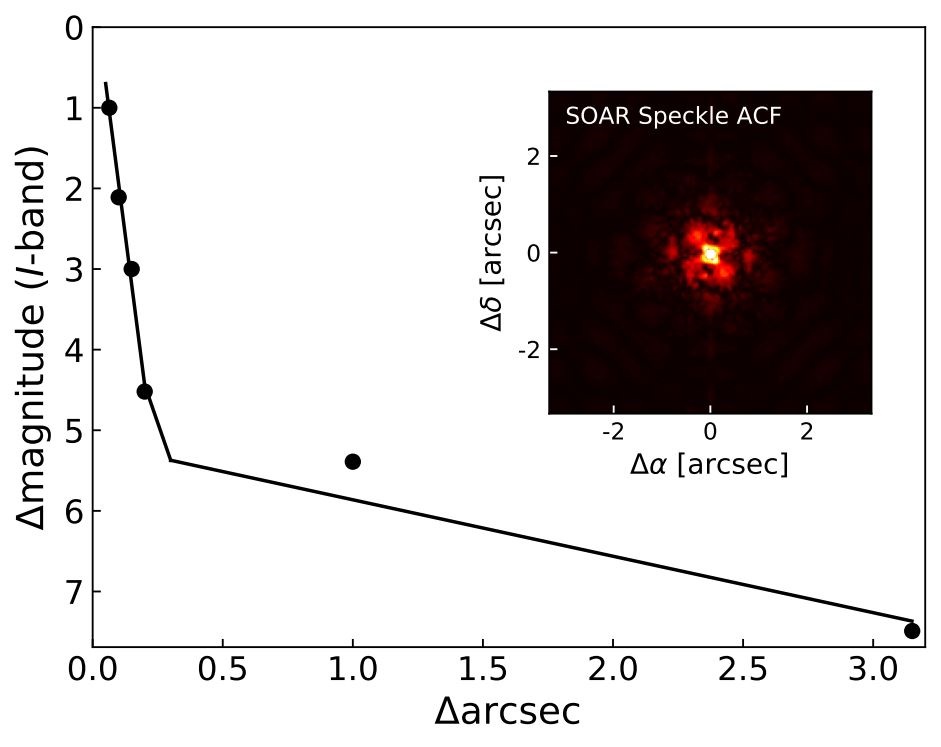}
\caption{The 5-$\sigma$ sensitivity limits and auto-correlation functions of the SOAR speckle observations of TOI-569. The black circles are measured data points and the lines are fits in two different separation regimes. In general, the sensitivity of speckle imaging to companions rises sharply from the diffraction limit to a ``knee" at a separation of $0.15 - 0.2\arcsec$, where it then continues to slowly increase out to ∼1.5\arcsec, beyond which the speckle patterns begin to become de-correlated. No nearby contaminating sources are detected within 3$\arcsec$.}\label{fig:soar}
\end{figure}

\dbf{Though the LCOGT data give us a sense of whether or not the transit signals for TOI-569 are within roughly 3$\arcsec$ of the target star, we may use speckle imaging to confirm whether or not there is contamination even closer to the target.} For TOI-569, we used SOAR speckle imaging to look for other objects within the TESS aperture that would significantly contaminate the transit and RV signals we observe. Nearby stars which fall within the same TESS image profile as the target can cause photometric contamination or be the source of an astrophysical false positive, such as a background or nearby eclipsing binary star. We searched for nearby sources to TOI-569 with SOAR speckle imaging \citep{tokovinin2018_soar} on May 18, 2019, observing in a similar visible bandpass as TESS (the Cousins-I band). Further details of the observations are available in \cite{ziegler2019_soar}. We detected no nearby stars within 3$\arcsec$ of TOI-569. The 5-$\sigma$ detection sensitivity and the speckle auto-correlation function from the SOAR observation are plotted in Figure~\ref{fig:soar}. 

We also use data from \textit{Gaia} DR2 \citep{gaiaDR2} to gather a census of nearby stars, finding that no stars brighter than G=17.0 are within 25$\arcsec$ and only two stars with G=13.7 and G=14.9 are approximately 26$\arcsec$ from TOI-569, which has a brightness of G=9.9 (Table \ref{tab:gaia}). These other fainter stars also do not share the same proper motion as TOI-569, which indicates that they are not associated with TOI-569 and are more distant background stars.

We do not have any high resolution imaging \dbf{(as in, sub-$3\arcsec$ coverage)} for TOI-1406, but using \textit{Gaia} DR2 data, we find only 3 other stars within 30$\arcsec$ of TOI-1406. The brightest of these other stars has a magnitude of G=15.8 and is 19$\arcsec$ from TOI-1406, which has a magnitude of G=11.8. We also find that none of these other stars share the same proper motion as TOI-1406 from the \textit{Gaia} DR2 data (Table \ref{tab:gaia}).

\subsection{CHIRON spectra}
\label{sec:chiron}
To characterize the RVs and stellar atmospheric parameters of TOI-569 and TOI-1406, we obtained a series of spectroscopic observations using the CHIRON spectrograph on the 1.5\,m SMARTS telescope \citep{chiron}, located at Cerro Tololo Inter-American Observatory, Chile. CHIRON is a high resolution echelle spectrograph that is fed via an image slicer and a fiber bundle. CHIRON achieves a spectral resolving power of $\lambda / \Delta \lambda \equiv R \sim 80,000$ over the wavelength region 4100 to $8700$\,\AA{}. The wavelength calibration is obtained via thorium-argon hollow-cathode lamp exposures that bracket each stellar spectrum.

To derive the stellar RVs, we performed a least-squares deconvolution \citep{1997MNRAS.291..658D} between the observed spectra and a non-rotating synthetic template generated via ATLAS9 atmospheric models \citep{Castelli:2004} at the stellar atmospheric parameters of each target. We then model the stellar line profiles derived from the least-squares deconvolution via an analytic rotational broadening kernel as per \citet{2005oasp.book.....G}. \dbf{This procedure follows that done in \cite{hats70b}.} The derived RVs for TOI-569 and TOI-1406 are listed in Table~\ref{tab:toi_rvs} \dbf{and this series of RVs helped to reveal the true orbital period of the BD.} The stellar parameters derived from the spectra of TOI-569 are $T_{\rm eff}=5669\pm 106$K, $\log{g}=4.11 \pm 0.18$, $\rm [Fe/H] = +0.23 \pm 0.05$ dex, and $v\sin{I_\star}=5.33\pm 0.50$ $\rm km\, s^{-1}$. \dbf{For TOI-1406, we find $T_{\rm eff}=6347\pm 186$K, $\log{g}=4.09 \pm 0.15$, $\rm [Fe/H] = -0.05 \pm 0.11$ dex, and an approximate full width at half max for the line broadening profile of $\rm FWHM=12.91\pm 0.24$ $\rm km\, s^{-1}$ with the CHIRON spectra. For TOI-569, we take care to account for the instrumental profile and macroturbulence to extract $v\sin{I_\star}$ from the FWHM approximation ($\rm FWHM = 6.65 \pm 0.5$ $\rm km\,s^{-1}$) with CHIRON as this is important in our analysis of the stellar inclination in Section \ref{sec:incl}.}

\begin{deluxetable*}{ccccccc}
\tabletypesize{\footnotesize}
\tablewidth{0pt}

 \tablecaption{Relative radial velocities, bisector span ($V_{\rm span}$), and FWHM of TOI-569 from CHIRON, CORALIE, and FEROS and of TOI-1406 from CHIRON and ANU. \label{tab:toi_rvs}}

 \tablehead{
 \colhead{$\rm BJD_{\rm TDB}-2450000$} & \colhead{RV ($\rm m\, s^{-1}$)} & \colhead{$\sigma_{\rm RV}$ ($\rm m\, s^{-1}$)} & \colhead{$V_{\rm span}$ ($\rm km\, s^{-1}$)} & \colhead{FWHM ($\rm km\, s^{-1}$)} & \colhead{Instrument} & \colhead{Target}}

\startdata
8594.61690 & 75132.3 & 16.6 & - & 6.93 & CHIRON & TOI-569\\
8606.50771 & 77754.5 & 20.7 & - & 6.67 & CHIRON & TOI-569\\
8596.60327 & 66265.0 & 18.9 & - & 7.00 & CHIRON & TOI-569\\
8595.59076 & 69864.3 & 16.4 & - & 6.67 & CHIRON & TOI-569\\
8607.53323 & 75982.6 & 17.8 & - & 6.73 & CHIRON & TOI-569\\
8611.59733 & 71966.8 & 24.9 & - & 6.65 & CHIRON & TOI-569\\
8612.57581 & 76679.3 & 30.6 & - & 6.57 & CHIRON & TOI-569\\
8649.44500 & 66193.1 & 24.2 & - & 6.63 & CHIRON & TOI-569\\
8651.48774 & 74957.8 & 34.9 & - & 6.81 & CHIRON & TOI-569\\
8654.46711 & 70453.9 & 27.2 & - & 6.40 & CHIRON & TOI-569\\
8593.58587 & 79347.5 & 13.0 & $-0.030$ & 10.55 & CORALIE & TOI-569\\
8597.47058 & 68533.2 & 31.6 & $0.089$ & 10.48 & CORALIE & TOI-569\\
8599.52446 & 78333.6 & 16.8 & $-0.023$ & 10.48 & CORALIE & TOI-569\\
8602.58567 & 69230.9 & 13.3 & $0.001$ & 10.46 & CORALIE & TOI-569\\
8603.46595 & 67533.0 & 20.3 & $-0.012$ & 10.47 & CORALIE & TOI-569\\
8614.49353 & 75527.3 & 32.4 & $0.004$ & 10.43 & CORALIE & TOI-569\\
8615.48815 & 70140.4 & 29.5 & $0.060$ & 10.66 & CORALIE & TOI-569\\
8594.49170 & 77148.3 & 6.5 & $0.019$ & - & FEROS & TOI-569\\
8595.52217 & 71681.8 & 8.4 & $0.015$ & - & FEROS & TOI-569\\
8597.51010 & 68686.8 & 6.8 & $0.015$ & - & FEROS & TOI-569\\  
8617.52568 & 70033.1 & 6.0 & $0.009$ & - & FEROS & TOI-569\\
\hline
8533.07797 & -19568.3 & 465.0 & - & - & ANU & TOI-1406\\
8534.98787 & -17679.4 & 197.8 & - & - & ANU & TOI-1406\\
8536.06364 & -15764.1 & 260.2 & - & - & ANU & TOI-1406\\
8537.96961 & -12090.1 & 725.7 & - & - & ANU & TOI-1406\\
8538.93516 & -12135.5 & 282.9 & - & - & ANU & TOI-1406\\
8561.89365 & -13798.5 & 270.0 & - & - & ANU & TOI-1406\\
8540.61381 & -13631.0 & 69.2 & - & 12.91 & CHIRON & TOI-1406\\
8541.60193 & -15950.3 & 91.8 & - & 12.69 & CHIRON & TOI-1406\\
8542.56709 & -17888.9 & 45.1 & - & 13.07 & CHIRON & TOI-1406\\
8544.52353 & -19087.2 & 166.4 & - & 12.78 & CHIRON & TOI-1406\\
8546.51779 & -15936.1 & 91.8 & - & 12.81 & CHIRON & TOI-1406\\
8562.57958 & -15177.2 & 126.9 & - & 12.89 & CHIRON & TOI-1406\\
8566.55925 & -18101.1 & 97.7 & - & 12.77 & CHIRON & TOI-1406\\
8567.59295 & -16029.1 & 83.6 & - & 12.82 & CHIRON & TOI-1406\\
8568.54388 & -13973.3 & 124.3 & - & 13.36 & CHIRON & TOI-1406\\
8569.57364 & -12432.3 & 110.3 & - & 12.98 & CHIRON & TOI-1406\\
\enddata
\end{deluxetable*}

\begin{figure}[!ht]
\centering
\includegraphics[width=0.40\textwidth, trim={0.0cm 0.0cm 1.5cm 0.0cm}]{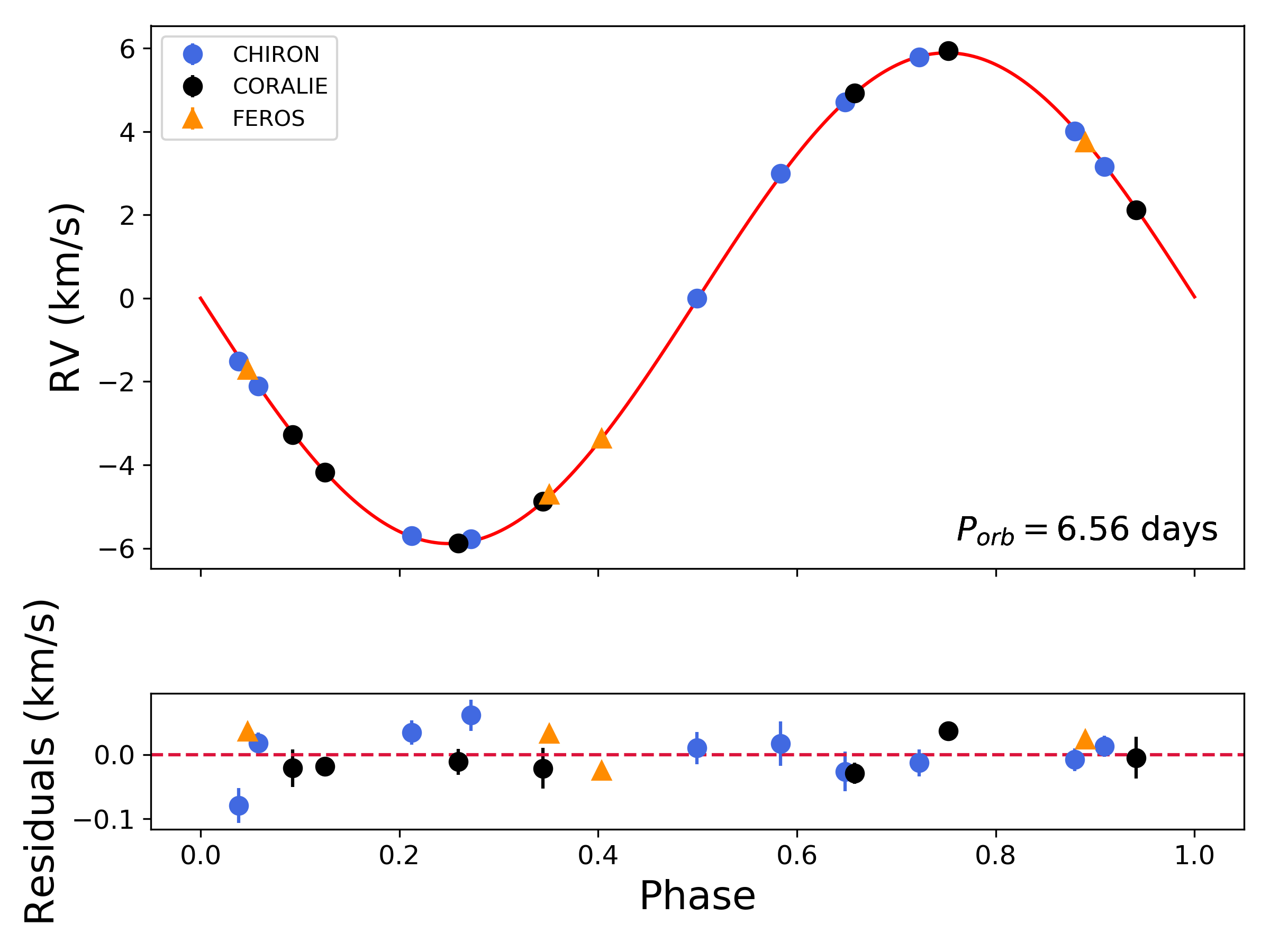}
\includegraphics[width=0.40\textwidth, trim={0.0cm 0.0cm 1.5cm 0.0cm}]{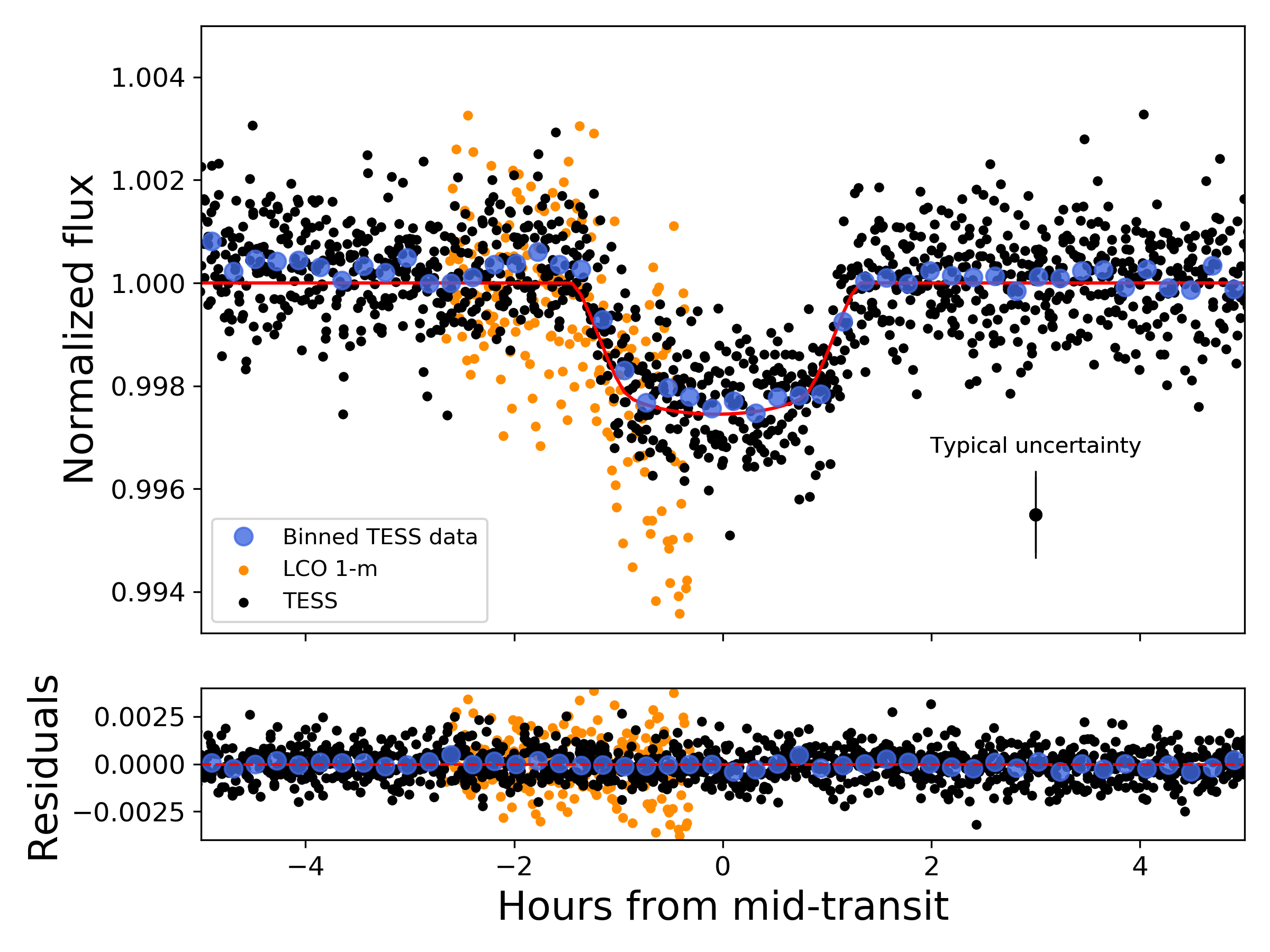}
\caption{{\it Top}: Relative radial velocities of TOI-569 with {\tt EXOFASTv2} orbital solution plotted in red. The orbital eccentricity is consistent with zero ($e < 0.0035$, 1-$\sigma$ upper limit). {\it Bottom}: TESS (black) and LCOGT 1-meter (orange) light curves with {\tt EXOFASTv2} transit model in red and binned TESS data in blue.}\label{fig:toi569_rv}
\end{figure}

\subsection{ANU 2.3m echelle spectra}
To help confirm TOI-1406b as a BD, we obtained six spectroscopic observations with the echelle spectrograph on the Australian National University (ANU) 2.3\,m telescope, located at Siding Spring Observatory, Australia. The ANU 2.3\,m echelle is a slit-fed spectrograph that yields a resolving power of $R \sim 23,000$ over the wavelength region of $\rm 3700-6700\,\AA$. Wavelength calibration was provided by bracketing thorium-argon lamp exposures, and the spectra were reduced as per \citet{2014MNRAS.437.2831Z}. The RVs from each exposure were measured via the least-squares deconvolution technique as described in Section~\ref{sec:chiron}. To derive $T_{\rm eff}$, $\log{g}$, and $\rm [Fe/H]$ for TOI-1406, we use {\tt SpecMatch-emp} \citep{yee2017_specmatch}, which matches the input spectra to a library of stars with well-determined parameters derived with a variety of independent methods, e.g., interferometry, optical and NIR photometry, asteroseismology, and LTE analysis of high-resolution optical spectra. From the ANU spectra and {\tt SpecMatch-emp} we find $T_{\rm eff}=6283 \pm 110$K, $\log{g}=4.13 \pm 0.12$, $\rm [Fe/H] = -0.09 \pm 0.09$ dex and $\rm   FWHM=15.0\pm 1.0$ $\rm km\, s^{-1}$ for TOI-1406.

\subsection{CORALIE spectra}
TOI-569 was observed with the CORALIE spectrograph on the Swiss 1.2\,m Euler telescope at La Silla Observatories, Chile \citep{CORALIE}, between April 19 and May 11, 2019. CORALIE has a resolving power of $R\sim 60,000$ and is fed by two fibers; one 2\arcsec\ diameter on-sky science fiber encompassing the star and another which can either be connected to a Fabry-P\'{e}rot etalon for simultaneous wavelength calibration or on-sky for background subtraction of the sky-flux. RVs were computed for each epoch by cross-correlating with a binary G2 mask \citep{Pepe2002}. Bisector-span, full-width half-max, and other line-profile diagnostics were computed as well using the standard CORALIE data reduction software. \dbf{Exposure times ranged from 450\,s to 1200\,s.} \dsout{Initial observations showed a 10.8 $\rm km\, s^{-1}$ RV shift over 3 days, and we subsequently reduced the exposure time from 1200 seconds to 450-600 seconds depending on seeing and airmass.} We obtain internal error estimates of 13-32 $\rm m\, s^{-1}$. The resulting velocities are plotted in Figure~\ref{fig:toi569_rv}, and are listed in Table~\ref{tab:toi_rvs}.

The CORALIE spectra were shifted to the stellar rest frame and stacked while weighting the contribution from each spectrum with its mean flux to produce a high signal-to-noise spectrum for spectral characterization using {\tt SpecMatch-emp} \citep{yee2017_specmatch}. We used the spectral region around the Mgb triplet ($5100 - 5340 {\rm \AA}$) to match our spectrum to the library spectra through $\chi^2$ minimization. A weighted linear combination of the five best matching spectra were used to extract bulk stellar parameters; $\teff=5481 \pm 110$K, $\log{g}=4.08 \pm 0.12$ and $\rm [Fe/H] = +0.41 \pm 0.09$ dex for TOI-569. 

\subsection{FEROS spectra}
TOI-569 was observed with the FEROS spectrograph \citep{feros} mounted on the MPG 2.2\,m telescope installed at the ESO La Silla Observatory. Four spectra were obtained between April 20 and May 14, 2019. Observations were performed with the simultaneous calibration mode where a second fiber is illuminated with a thorium-argon lamp for tracking the instrumental drift in RV during the science exposure. The adopted exposure time was of 400s which produced spectra with a typical signal-to-noise ratio per resolution element of 90. FEROS data were processed with the {\tt ceres} pipeline \citep{ceres}, which performs the optimal extraction of the raw data, the wavelength calibration, the instrumental drift correction, and the computation of precise RVs and bisector spans. The results are presented in Table~\ref{tab:toi_rvs}. The four FEROS spectra were combined in order to measure the atmospheric parameters using the \texttt{zaspe} package \citep{zaspe}, obtaining $T_{\rm eff}$ = 5669 $\pm$ 80 K, $\log{g}$ = 4.21 $\pm$ 0.12, $\rm [Fe/H] = +0.28 \pm 0.05$ dex, and an approximate $\rm   FWHM = 6.45 \pm 0.30$ $\rm km\, s^{-1}$ for TOI-569.

\begin{figure}[!ht]
\centering
\includegraphics[width=0.40\textwidth, trim={0.0cm 0.0cm 1.5cm 0.0cm}]{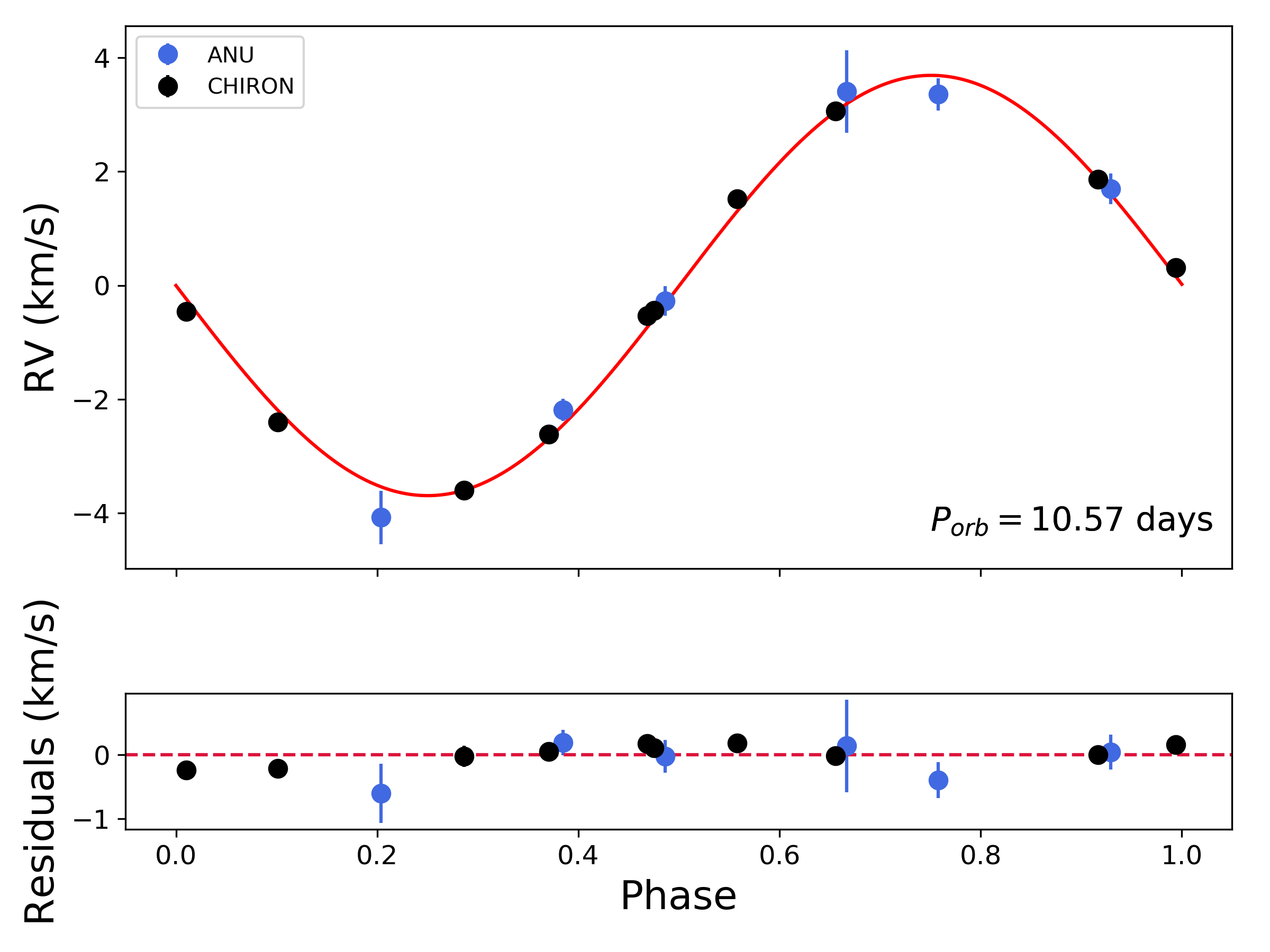}
\includegraphics[width=0.40\textwidth, trim={0.0cm 0.0cm 1.5cm 0.0cm}]{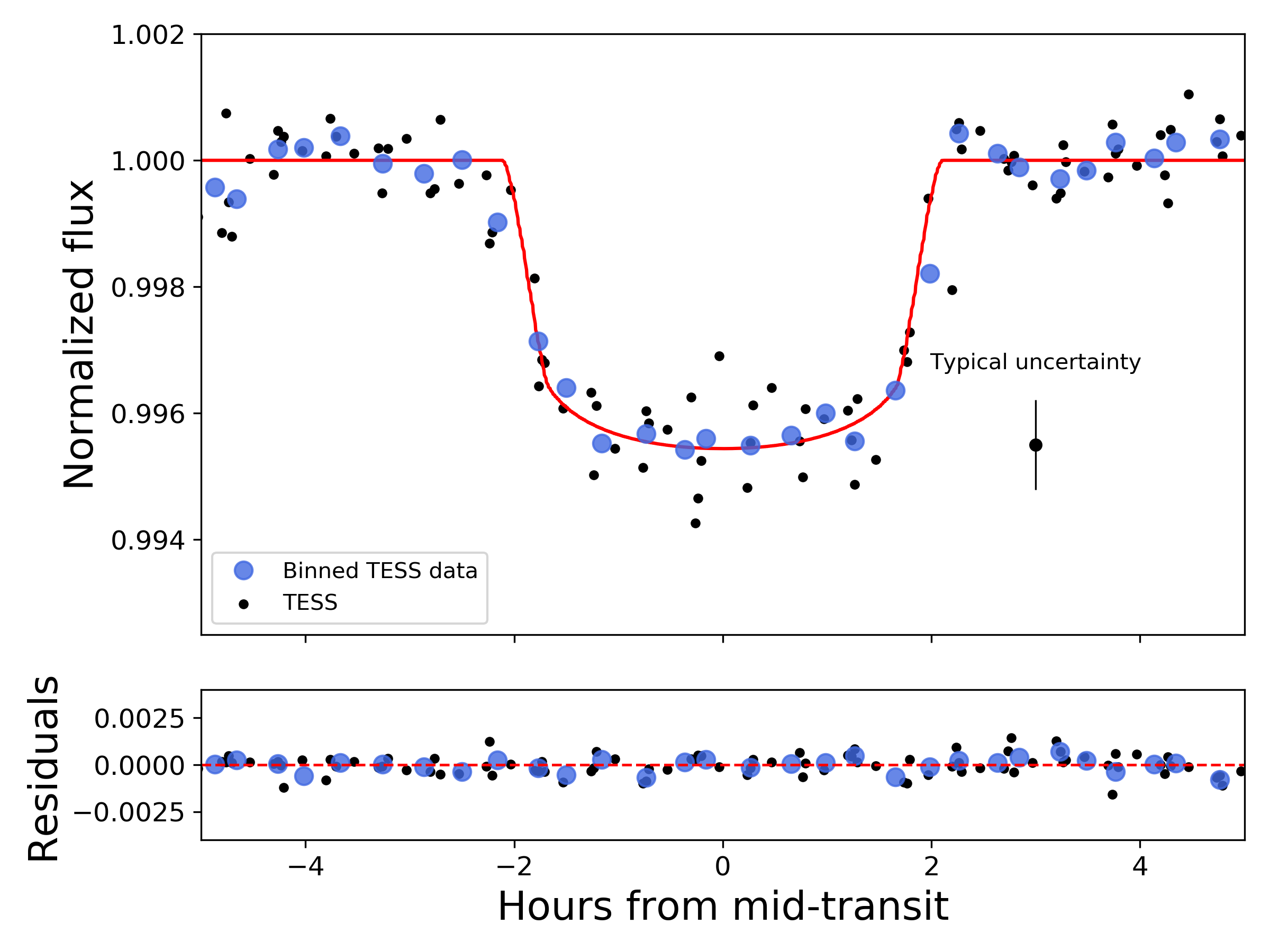}
\caption{{\it Top}: Relative radial velocities of TOI-1406 with {\tt EXOFASTv2} orbital solution plotted in red. The orbital eccentricity is consistent with zero ($e < 0.039$, 1-$\sigma$ upper limit). {\it Bottom}: TESS light curve with {\tt EXOFASTv2} transit model in red and binned TESS data in blue.}\label{fig:toi1406_rv}
\end{figure}

\begin{deluxetable}{cccc}
\tabletypesize{\footnotesize}
\tablewidth{0pt}

 \tablecaption{Spectroscopic values for TOI-569 and TOI-1406 from CHIRON, ANU, CORALIE, and FEROS. We use [Fe/H] and $T_{\rm eff}$ values from CHIRON as inputs to the global model described in Section \ref{sec:exofast}.\label{tab:spec_values}}

 \tablehead{
 \colhead{TOI-569} & \colhead{CHIRON} & \colhead{CORALIE} & \colhead{FEROS} }

\startdata 
$T_{\rm eff}$ (K) &  $5669 \pm 106$ & $5481 \pm 110$ & $5669 \pm 80$ \\
$\log{g}$ &  $4.11 \pm 0.18$ & $4.08 \pm 0.12$ & $4.21 \pm 0.12$ \\
$\rm [Fe/H]$ (dex) &  $+0.23 \pm 0.05$ & $+0.41 \pm 0.09$ & $+0.28 \pm 0.05$ \\
FWHM ($\rm km\,s^{-1}$) &  $6.65 \pm 0.16$ & $10.50 \pm 0.50$ & $6.45 \pm 0.30$ \\
$R$ (resolution) & 80,000 & 60,000 & 48,000 \\
\hline
TOI-1406 & ANU & CHIRON & \\
\hline
$T_{\rm eff}$ (K) & $6283 \pm 110$ & $6347 \pm 186$ & -  \\
$\log{g}$ & $4.13 \pm 0.12$ & $4.09 \pm 0.15$ & - \\
$\rm [Fe/H]$ (dex) & $-0.09 \pm 0.09$ & $-0.05 \pm 0.11$ & -  \\
FWHM ($\rm km\,s^{-1}$) & $15.0 \pm 1.0$ & $12.9 \pm 0.2$ & - \\
$R$ (resolution) & 23,000 & 80,000 & - \\
\enddata
\end{deluxetable}

\section{Analysis} \label{sec:analysis}
\subsection{Modeling with {\tt EXOFASTv2}}\label{sec:exofast}
The masses and radii of the BDs are derived using {\tt EXOFASTv2}. A full description of {\tt EXOFASTv2} is given in \cite{eastman2019}. {\tt EXOFASTv2} uses the Monte Carlo-Markov Chain (MCMC) method. For each MCMC fit, we use N=36 (N = 2$\times n_{\rm parameters}$) walkers, or chains, and run for 50,000 steps, or links. To derive stellar parameters, {\tt EXOFASTv2} utilizes the MIST isochrone models \citep{mist1, mist2, mist3}. \dsout{or the Yonsei-Yale isochrone models. We use the MIST models for the analysis of TOI-1406 and the YY models for TOI-569. The YY models include a metallicity range of up to $\rm [Fe/H] = +0.78$ dex while the MIST models currently implemented in {\tt EXOFASTv2} extend [Fe/H] to only $+0.5$ dex. Based on our spectroscopic measurements of the metallicity for TOI-569 (roughly $+0.4$ dex), we choose to report the results for TOI-569 from the YY models in order to avoid approaching the $+0.5$ dex boundary on metallicity in the MIST models during the MCMC analysis. We see a difference of 0.1 dex between the metallicity found by the MIST models ($\rm [Fe/H]_{MIST} = +0.30^{+0.14}_{-0.21}$ dex) to that found by the YY models ($\rm [Fe/H]_{YY} = +0.40^{+0.07}_{-0.08}$ dex).} 

The parameters for which we set priors and the types of priors we set for each (i.e. uniform $\mathcal{U}[a,b]$ or Gaussian $\mathcal{G}[a,b]$) are shown in Tables \ref{tab:exofast_toi1406} and \ref{tab:exofast_toi569}. We rely on our spectroscopic measurements \dbf{of [Fe/H] and $T_{\rm eff}$} and parallax measurements from \textit{Gaia} to define our Gaussian priors, which penalize the fit for straying beyond the width, $b$, away from the mean, $a$ of the parameter. We use an upper limit for the $A_V$ extinction. See Table 3 of \cite{eastman2019} for a detailed description of priors in {\tt EXOFASTv2}. \dbf{For the choice of priors for [Fe/H] and $T_{\rm eff}$, we use the CHIRON values since CHIRON has the highest spectral resolution $R = 80,000$ of the spectrographs we used (see Table \ref{tab:spec_values}). The resulting {\tt EXOFASTv2 values are consistent with the input values from CHIRON.}} The spectral energy distribution for each star is also taken into account with {\tt EXOFASTv2} and are shown in Figure \ref{fig:sed}. \dbf{The BD parameters are derived with the normalized TESS and LCOGT light curves and non-phase folded RVs into {\tt EXOFASTv2} as inputs. The non-negligible BD mass is properly accounted for in {\tt EXOFASTv2}, so no particularly special treatment is needed with regard to deriving the companion masses.}

\begin{figure}[!ht]
\centering
\includegraphics[width=0.45\textwidth, trim={1.0cm 0.0cm 0.0cm 0.0cm}]{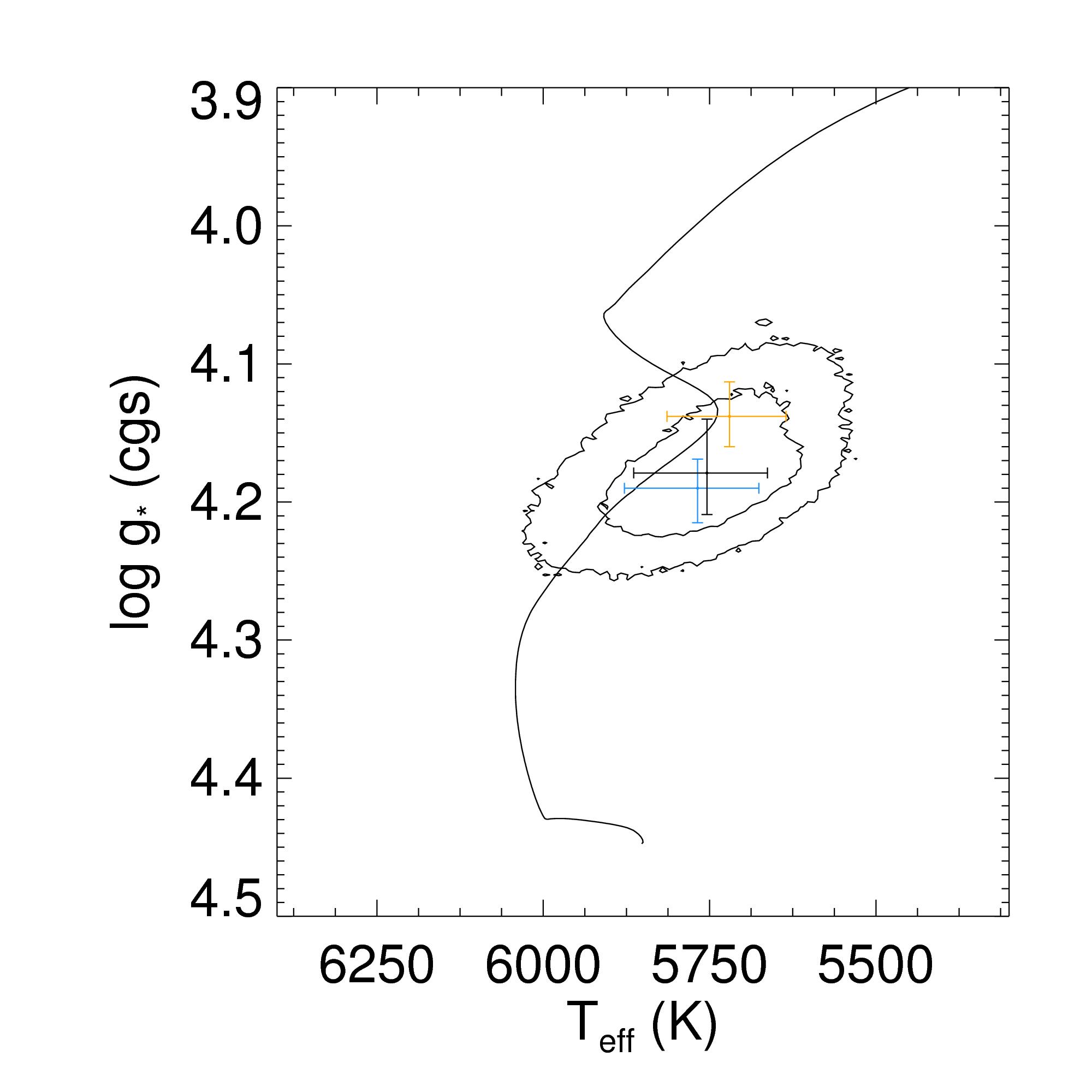}
\includegraphics[width=0.45\textwidth, trim={0.0cm 0.0cm 0.0cm 0.0cm}]{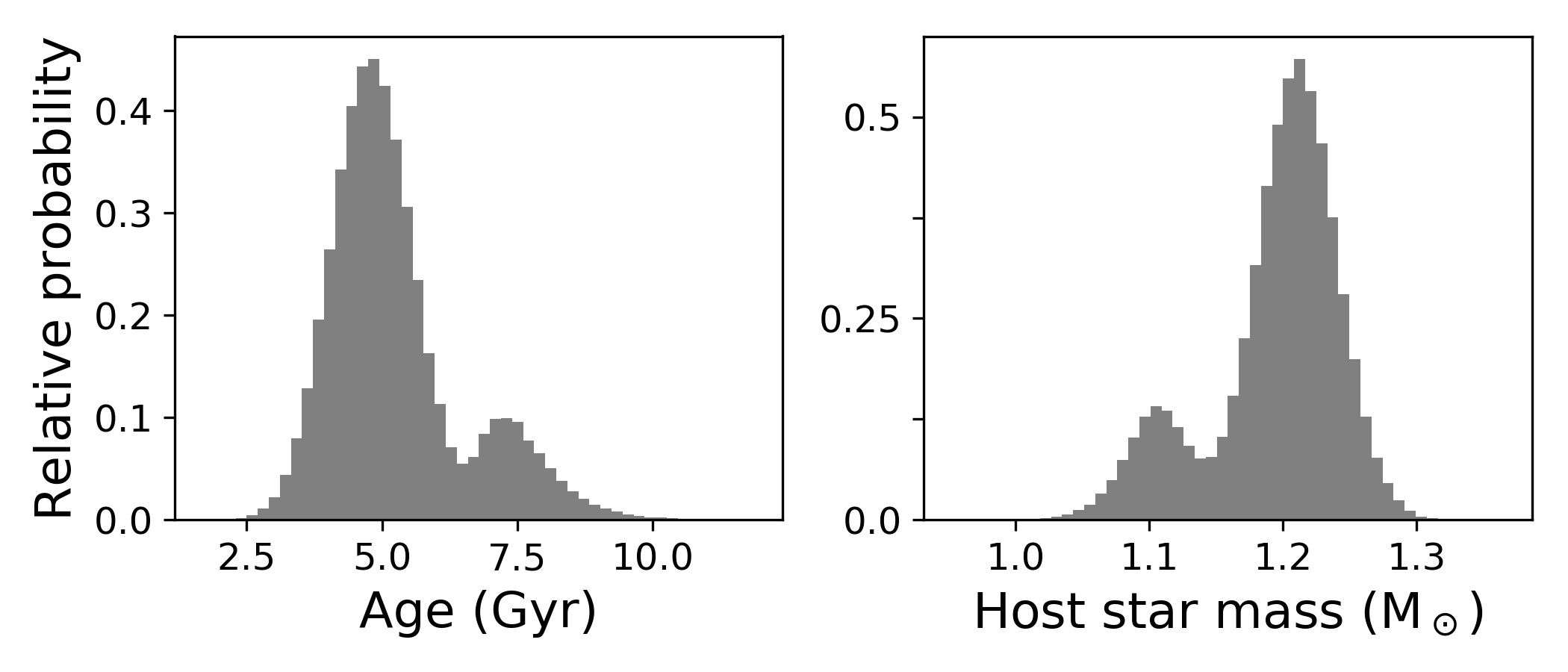}
\caption{{\it Top}: \dbf{MIST} isochrone from {\tt EXOFASTv2} for TOI-569. 
\dbf{The best-fitting \dbf{MIST} track is shown by the black line. The median values and 1-$\sigma$ errors from our global fit are shown in the black cross with the corresponding 3-$\sigma$ contours in black. When splitting this bimodal global solution (black points and contours), the results are the blue and orange crosses. The blue cross shows the higher probability solution for $\log{g}$ and $T_{\rm eff}$ and the orange cross shows the lower probability solution.} \dsout{The bimodality of the {\tt EXOFASTv2} solution arises because TOI-569 falls near the main sequence turnoff point.} {\it Bottom}: Age and stellar mass posterior distributions from {\tt EXOFASTv2} for TOI-569. We show these to provide a sense of the relative probabilities between the peaks of the bimodal distributions, which is roughly 3-to-1 in favor of a more massive, younger system (blue cross in top panel). We see no bimodality for the posterior distributions of TOI-1406 in {\tt EXOFASTv2}. \label{fig:bimodal}}
\end{figure}

We see bimodality in the posterior distribution for the age (and correlated parameters) of TOI-569, so we present the two most probable solutions resulting from the bimodal posterior distributions with the absolute most probable solution taken as the final \dsout{reported} \dbf{adopted} value (Table~\ref{tab:exofast_toi569}). The most relevant bimodal posterior distributions are shown in Figure \ref{fig:bimodal}. The probability of the solution, for age and mass, we report here is 0.73, with the less likely solution having a probability of 0.27.

The resulting stellar SED models from {\tt EXOFASTv2} for TOI-569 and TOI-1406 are shown in Figure \ref{fig:sed}. These follow the procedures outlined in \citet{Stassun:2016,Stassun:2017,Stassun:2018}.

\begin{figure}[!ht]
\centering
\includegraphics[width=0.45\textwidth, trim={0.0cm 0.0cm 0.0cm 0.0cm}]{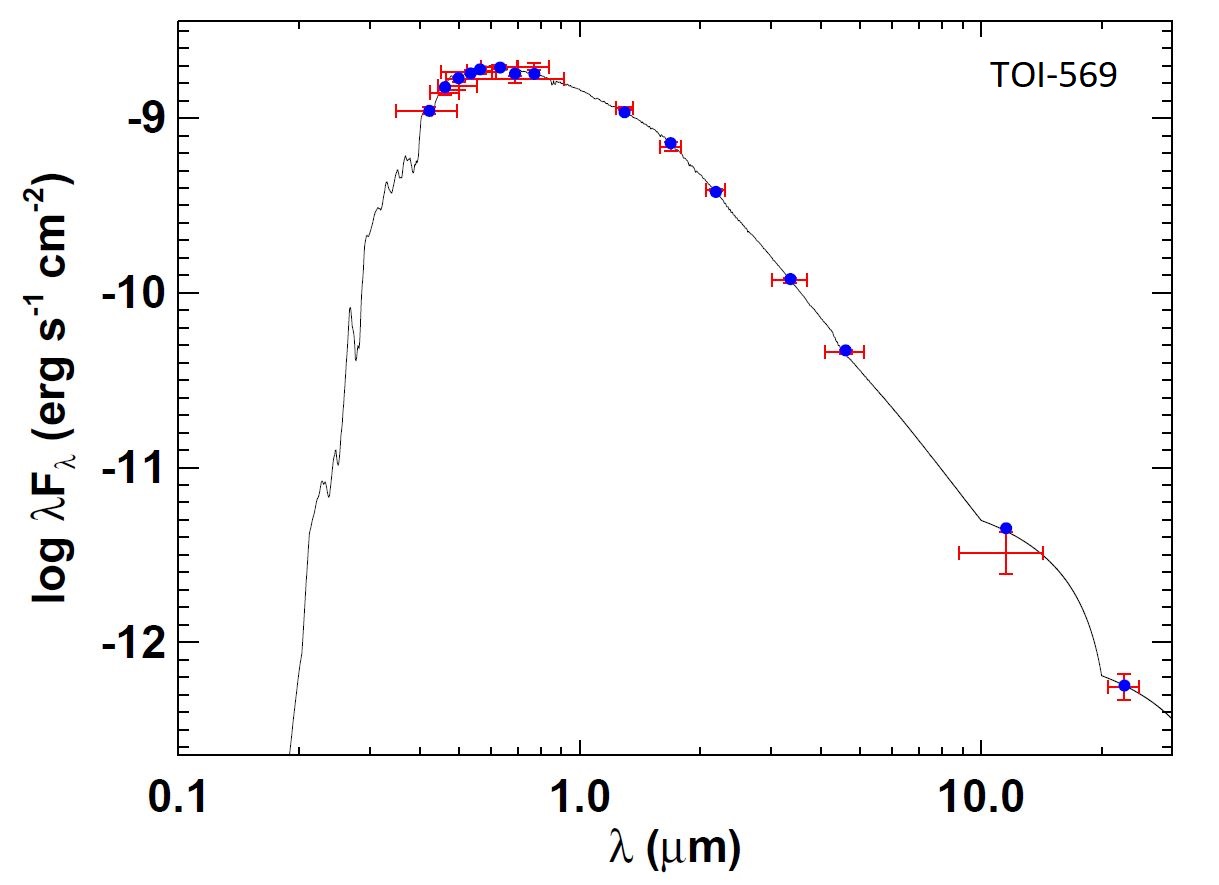}
\includegraphics[width=0.45\textwidth, trim={0.0cm 0.0cm 0.0cm 0.0cm}]{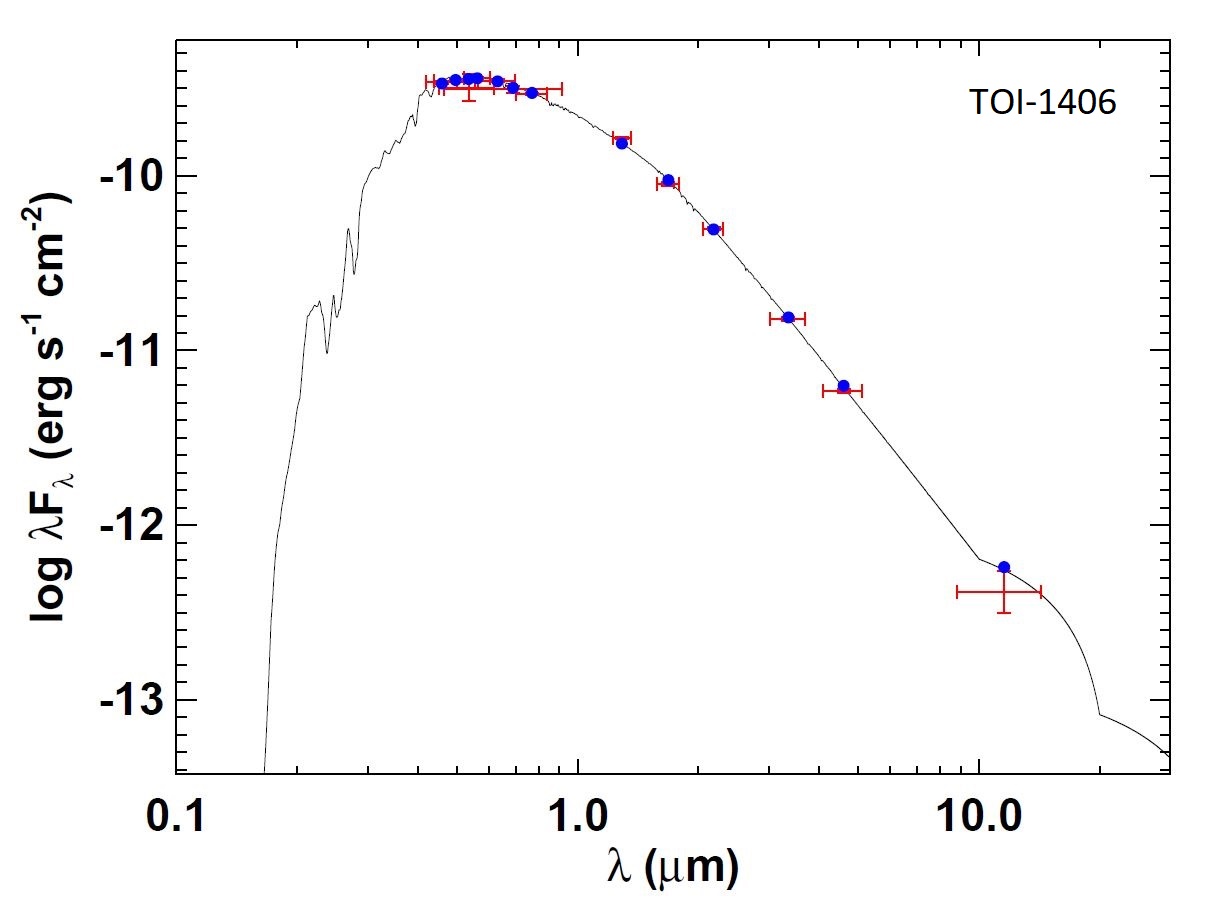}
\caption{Spectral energy distributions for TOI-569 and TOI-1406. Red symbols represent the observed photometric measurements, where the horizontal bars represent the effective width of the bandpass. Blue symbols are the model fluxes from the best-fit Kurucz atmosphere model (black). \label{fig:sed}}
\end{figure}

\subsection{Analysis with {\rm \texttt{pyaneti}}}
As an independent check on our {\tt EXOFASTv2} analysis, we also carried out an analysis with the {\tt{pyaneti}}\footnote{\url{https://github.com/oscaribv/pyaneti}} \citep{2019MNRAS.482.1017B} software. Using a Bayesian approach combined with MCMC sampling, we performed a joint analysis of the RV measurements and the TESS light curves and modelled posterior distributions of the fitted parameters. The RV data were fitted with Keplerian orbits, and for each different instrumental set-up, an offset term for each systemic velocity is included. \dbf{The non-negligible mass of the brown dwarf is properly taken into account in {\tt pyaneti} \citep{barragan2019}.} The photometric data are modeled with the quadratic limb-darkening model of \citet{managol02}.

We use uniform priors and fit for the BD-to-star radius ratio, the orbital period, the mid-transit time, the scaled orbital distance, the eccentricity, the argument of periastron, the impact parameter \mbox{($b$}), and the Doppler semi-amplitude variation ($K$). The allowed ranges for the fit parameters for {\tt pyaneti} are shown in Table~\ref{tab:pyaneti}.

\begin{deluxetable}{ccc}
\tabletypesize{\footnotesize}
\tablewidth{0pt}

 \tablecaption{Allowed ranges for fit parameters from {\tt pyaneti}.\label{tab:pyaneti}}

 \tablehead{
 \colhead{Parameter} & \colhead{TOI-569} & \colhead{TOI-1406}}

\startdata 
$R_{\rm BD}/R_\star$ & [0, 0.1] & [0, 0.1] \\
$P_{\rm orb}$ (days) & [6.5541, 6.5580] & [10.5721, 10.5762] \\
$T_0-2458400$ ($\rm BJD_{TDB}$) & [96.858, 96.878] & [14.5061, 14.7061] \\
$a/R_\star$ & [1.1, 12] & [1.1, 19]\\
$e\cos{\omega}$ & [-1, 1] & [-1, 1] \\
$e\sin{\omega}$ & [-1, 1] & [-1, 1] \\
Impact parameter $b$ & [0, 1] & [0, 1] \\
Semi-amplitude $K$ ($\rm km\, s^{-1}$) & [0, 15] & [0, 15] \\
\enddata
\end{deluxetable}

We used 500 independent chains, and checked for convergence after every 5,000 iterations. After convergence, a posterior distribution of 250,000 independent points for every parameter was computed from the last 500 iterations. We find the eccentricity to be consistent with zero for both BDs. We find a mass and radius of TOI-569b and TOI-1406b to be consistent within 1-$\sigma$ of the values from the {\tt EXOFASTv2} models.

\subsection{Rotational inclination angle of TOI-569}\label{sec:incl}
\begin{figure}[!ht]
\centering
\includegraphics[width=0.49\textwidth, trim={0.0cm 0.0cm 0.0cm 0.0cm}]{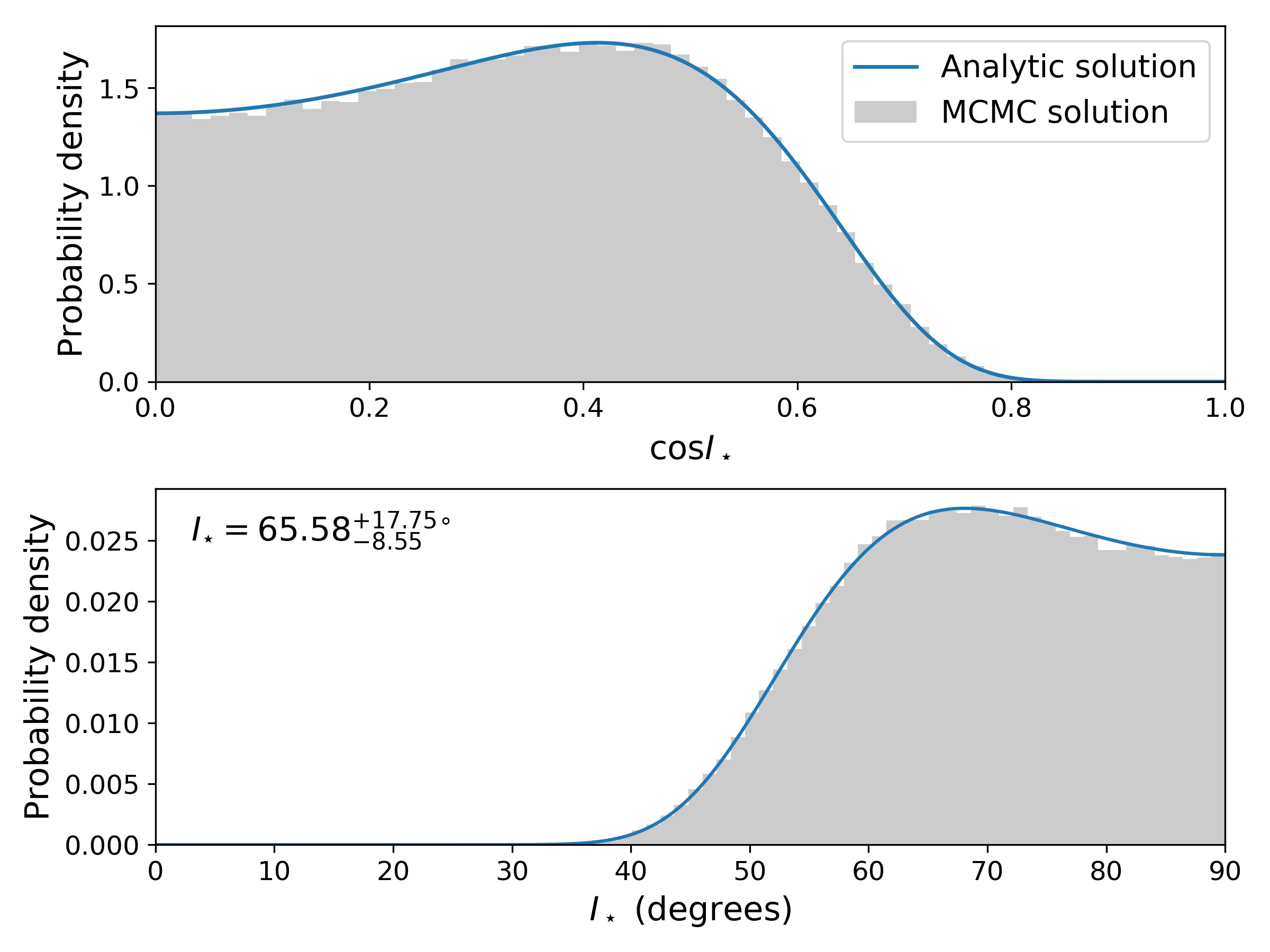}
\caption{Probability distributions of $\cos{I_\star}$ and $I_\star$ for TOI-569. The analytic and MCMC solutions follow the procedure outlined by \cite{masuda2020}. \label{fig:stellar_incl}}
\end{figure}

\dbf{Astronomers may calculate the angle at which a star is inclined to the line-of-sight, $I_\star$, in order to learn about the relative alignment between this angle and the orbital inclination angle, $i$, of a transiting or eclipsing object. Using the spectroscopic $v\sin{I_\star}$ measurement from CHIRON (the highest resolution spectrograph we used with $R \sim 80,000$) and the $P_{\rm rot}$ results from the Lomb-Scargle periodogram analysis, we calculate the inclination of the rotational axis of TOI-569 to be $I_\star=65.58^{+17.75}_{-8.55}$\,$^\circ$ (1-$\sigma$ uncertainties). This is traditionally done by taking: 
\begin{equation}\label{eq:inclination}
    I_\star = \sin^{-1}\left(\frac{v\sin{I_\star}}{V_{\rm rot}}\right)
\end{equation}
where $v\sin{I_\star}=5.33 \pm 0.50$ $\rm km\, s^{-1}$ and $V_{\rm rot} = 2\pi R_\star/P_{\rm rot} = 5.80 \pm 0.58$ $\rm km\, s^{-1}$. However, this traditional technique neglects the dependence of the priors on $v\sin{I_\star}$ and $V_{\rm rot}$ on each other. \cite{masuda2020} provide guidance on how to properly address this flaw with the traditional technique. We show our results of $I_\star$ for TOI-569 using the formulation described by \cite{masuda2020} in Figure \ref{fig:stellar_incl}. We use the MCMC distribution to calculate the 1-$\sigma$ uncertainties as the $16^{\rm th}$ and $84^{\rm th}$ percentiles of the distribution with the mean being the peak of the analytic distribution. We use the peak of the distribution because the distribution is skewed and so the median would bias $I_\star$ to higher values. This method neglects the affects of differential rotation in the star, which may change the $P_{\rm rot}$ we report here by up to 15\% depending on the latitude of the star spots \citep{quinn2016}.}

\dbf{The orbital inclination of TOI-569b is $i = 85.37^{+0.13}_{-0.11}\,^\circ$. Given the probability distribution of the stellar inclination $I_\star = 65.58^{+17.75}_{-8.55}\,^\circ$ of TOI-569 (see Figure \ref{fig:stellar_incl}), we argue that this system is marginally misaligned and that alignment cannot be ruled out.}

\dbf{When only $v\sin{I_\star}$ is known, as is the case with TOI-1406, we use the following equation to place an upper limit on $P_{\rm rot}$ (5.3 days), which is much shorter than the orbital period of TOI-1406b (10.6 days), meaning that the system is not synchronized}:
\begin{equation}\label{eq:prot}
    P_{\rm rot} \leq \frac{2\pi R_\star}{v\sin{I_\star}}
\end{equation}

\subsection{Tidal circularization timescales}\label{subsec:circularization}
\dbf{Over time, tidal interactions between a host star and any companions affect their orbits. Generally, the orbits of the companions and their host stars first begin to circularize according to what is known as the circularization timescale. Next, the orbital period of the companion synchronizes with their host star's rotation (the synchronization timescale). Finally, the system experiences a spin-orbit co-alignment \citep{mazeh08}. These timescales are influenced by the mass, radius, separation, and tidal quality factor $Q$ of both the host star and companion of a system.} \dsout{In the case of EPIC 212036875b, a short-period BD with mass $M_b=51 \pm 2\mj$, Persson et al. (2019) found that the BD still has some orbital eccentricity ($e=0.132$) and its current orbital configuration is a result of a relatively quick inward migration, favoring a formation scenario that the BD did not form at or near its current orbital configuration (i.e., not in-situ). In the case of TOI-503b, \v{S}ubjak et al. 2019 find a relatively young ($\sim$180 Myr) A-type host star with a BD ($M_b=53.7\mj$) in a circular orbit. This system has been more complex to describe with established tidal interaction models (Jackson et al. 2008) given that these models are not tailored to \dsout{A-type} stars with radiative envelopes and BD companions as massive as TOI-503b. Given this, \v{S}ubjack et al. 2019 do not conclusively determine a formation scenario for TOI-503b (in-situ versus not) and instead offer some discussion of the tidal quality factors, $Q$, that may be most appropriate for this system. Here, we will examine the circularization timescales for TOI-569 and TOI-1406 in a similar way.} \dbf{Here, we restrict our discussion to the circularization timescales for different values of the tidal quality factors, $Q_\star$ and $Q_{\rm BD}$, that may be most appropriate for the TOI-569 and TOI-1406 systems.}

Following the formalism from \cite{Jackson2008}, the equations for the orbital circularization timescale for a close-in companion are:
\begin{equation}\label{eq:tides_star}
    \frac{1}{\tau_{\rm circ,\star}} = \frac{171}{16}\sqrt{\frac{G}{M_\star}}\frac{R_\star^5M_{\rm BD}}{Q_\star}a^{-\frac{13}{2}}
\end{equation}

\begin{equation}\label{eq:tides_planet}
    \frac{1}{\tau_{\rm circ, BD}} = \frac{63}{4}\frac{\sqrt{GM_\star^3}R_{\rm BD}^5}{Q_{\rm BD}M_{\rm BD}}a^{-\frac{13}{2}}
\end{equation}



\begin{equation}\label{eq:tides_e}
    \frac{1}{\tau_{\rm e}}= \frac{1}{\tau_{\rm circ, \star}} + \frac{1}{\tau_{\rm circ, BD}}
\end{equation}

\begin{deluxetable}{cccc}
\tabletypesize{\footnotesize}
\tablewidth{0pt}

 \tablecaption{Circularization timescales for different values of $Q_\star$ and $Q_{\rm BD}$ with stellar rotational period and BD orbital period also shown. \dbf{We quote the rotation period measured from the periodogram of the TESS light curve for TOI-569.} \dbf{An upper limit on the rotation period of TOI-1406 is calculated (Equation \ref{eq:prot}) using $R_\star$ from {\tt EXOFASTv2} and FWHM as an approximation for $v\sin{I_\star}$}. \dbf{Since we do not directly have the inclination of the star's rotation axis for TOI-1406, the rotation period listed for this star is an upper limit. The $v\sin{I_\star}$ and FWHM values are from CHIRON.} \dsout{Note for TOI-1406, the shortest $\tau_{\rm e}$ for reasonable choices of $Q_\star=10^7$ and $Q_{\rm BD}=10^{4.5}$ is $\tau_{\rm e}=91.5^{+18.4}_{-17.0}$ Gyr.} These show 1-$\sigma$ uncertainties. \label{tab:circularization}}

 \tablehead{
 \colhead{Object name \& Age} & \colhead{$Q_\star$} & \colhead{$Q_{\rm BD}$} & \colhead{$\tau_{\rm e}$ (Gyr)}}

\startdata 
TOI-569 & $10^{7}$ & $10^{6}$ & $8.0^{+0.8}_{-1.2}$\\
 $4.70^{+1.30}_{-1.30}$ Gyr  & $10^{7}$ & $10^{4.5}$ & $7.5^{+0.7}_{-1.1}$\\
 & $10^{6}$ & $10^{4.5}$ & $0.80^{+0.08}_{-0.12}$\\
 & $10^{5}$ & $10^{4.5}$ & $0.08^{+0.01}_{-0.01}$\\
 \hline
TOI-1406 & $10^{7}$ & $10^{6}$ & $127^{+26.2}_{-24.1}$\\
 $3.20^{+2.20}_{-1.60}$ Gyr &  $10^{7}$ & $10^{4.5}$ & $91.5^{+18.4}_{-17.0}$\\
 & $10^{6}$ & $10^{4.5}$ & $12.4^{+2.5}_{-2.3}$\\
 & $10^{5}$ & $10^{4.5}$ & $1.3^{+0.3}_{-0.2}$\\
 \hline
 & $P_{\rm rot}$ (days) & $P_{\rm orb}$ (days) & $v\sin{I_\star}$ ($\rm km\, s^{-1}$) \\
 \hline
 TOI-569  & \dbf{12.9} & $6.6$ & $  5.3 \pm 0.5$\\
 \hline
 & $P_{\rm rot}$ (days) & $P_{\rm orb}$ (days) & $\rm FWHM$ ($\rm km\, s^{-1}$) \\
 \hline
 TOI-1406 & \dbf{$\leq 5.3$} & $10.6$ & $  12.9 \pm 0.2$\\
\enddata
\end{deluxetable}

\noindent where $\tau_{\rm e}$ is the circularization timescale, $a$ is the semi-major axis, $M_\star$ is the stellar mass, $R_\star$ is the stellar radius, $M_{\rm BD}$ is the BD mass, $R_{\rm BD}$ is the BD radius, $Q_\star$ is the tidal quality factor for the star, and $Q_{\rm BD}$ is the tidal quality factor for the BD. \dsout{The individual contributions of the effects of tides raised on the star and the BD are accounted for in Equations \ref{eq:tides_star} \& \ref{eq:tides_planet}, respectively, and summed together in Equation \ref{eq:tides_e}.} Equation \ref{eq:tides_e} is a prediction on how long it takes for the orbital eccentricity of an object to decrease by an exponential factor (the relationship $\tau_{\rm e} \propto dt \propto -de/e$) \dbf{based on the tides raised on the star and BD}. 

Use of this equation comes with a number of assumptions that we reiterate here from \cite{Jackson2008}: 1) the BD is in a short orbital period (10 days or less), 2) the orbital eccentricity $e$ is small (though for companions in the planetary mass range, higher-order terms may be important to account for higher $e$ in the past), 3) the BD's orbital period $P_{\rm orb}$ is smaller than the host star's rotation period $P_{\rm rot}$, \dbf{and 4) $Q_\star$ is independent of the tidal forcing frequency}. Admittedly, Equation \ref{eq:tides_e} and these assumptions cater to hot Jupiters and not the type of more massive BDs in this study. \dsout{which is important to highlight as $P_{\rm rot}$ of the host star is influenced by the presence of a massive companion. We also note that Equation \ref{eq:tides_e} produces longer estimates for the circularization timescale compared to solutions worked out by Zahn 1977 and others, which predict tidal evolution to occur on shorter timescales.}

\dsout{The circularization timescale and the synchronization timescale, $\tau_{\Omega}$, change over time as the host star and especially the BD change in radius. The equation for the synchronization timescale, $\tau_{\Omega}$, from Goldreich 1966 is:}
\dsout{where $\Omega_\star$ is the angular velocity of the stellar rotation and $\alpha_\star$ is the stellar moment of inertia.}

With these considerations in mind, we calculate $\tau_{\rm e}$ for TOI-569b and TOI-1406b for a range of $Q_\star$ and $Q_{\rm BD}$ for each system in Table~\ref{tab:circularization}. \dsout{The rotation period of while rotation period for TOI-1406 is estimated using the projected rotation rate $v\sin{I_\star}$ and the radius of the star derived from {\tt EXOFASTv2}.} The choice to adopt a $Q_{\rm BD}$ as low as $10^{4.5}$ comes from \cite{cww89a}, who directly constrain $Q_{\rm BD}$ for CWW 89Ab, a $M_{\rm BD}=39\mj$ BD. \dbf{The choice to adopt a $Q_\star$ as low as $10^5$ comes from studies of circularization of binary stars \citep{meibom05, milliman14}.} For the bimodal posterior distributions of TOI-569, we only use the most probable $M_\star$, $R_\star$, $M_{\rm BD}$, $R_{\rm BD}$, and $a$ (Table~\ref{tab:exofast_toi569}). \dsout{Each parameter ($M_\star$, $R_\star$, $M_{\rm BD}$, $R_{\rm BD}$, and $a$) is a randomly sampled posterior distribution from {\tt EXOFASTv2} and so the $\tau_{\rm e}$ distribution is as well. From the $\tau_{\rm e}$ distribution for our selected combinations of $Q_\star$ and $Q_{\rm BD}$, we calculate the median and 1-$\sigma$ uncertainties as the $16^{\rm th}$ and $84^{\rm th}$ percentiles of the distribution.} \dsout{For TOI-569, we find a range of $\tau_{\rm e} = 7.5$ Gyr to $\tau_{\rm e} = 78.6$ Gyr. For TOI-1406, we see $\tau_{\rm e} > 91$ Gyr for the lowest reasonable choices of tidal quality factors, however, since $P_{\rm orb} > P_{\rm rot}$ in this system, these timescales are generally invalid. We also note that Jackson et al. 2008 provide equations for the orbital decay timescale $\tau_{a}$ and for both TOI-569 and TOI-1406, the decay timescale is at least an order of magnitude longer than the age of either system.}

\dbf{We highlight the tidal theory here to show that for these BDs, it is difficult to pin down the timescale over which tidal interactions influence their orbits. Though both BDs have circular orbits, we may only conclude that TOI-1406b likely underwent a low-eccentricity migration unless tidal dissipation was extremely efficient. The circularization timescales for TOI-569 may be short enough such that tidal interactions alone would have circularized the orbit of the BD over the system's age, thus making it difficult to tell whether or not the BD formed in a circular orbit.}

\section{Discussion}\label{sec:conclusion}
Including the two new BDs in this work, the total number of known BDs that transit a star is 23 (Table~\ref{tab:bdlist}). With the discovery of TOI-569b and TOI-1406b, the total number of new transiting BDs discovered or observed by the TESS mission is now 4 \citep[][this work]{subjak2019, jackman2019}. We expect at least as many more to be discovered as TESS continues its observations over the remainder of its primary mission. At present, we do not have enough transiting BDs to perform a statistical study of the population and draw conclusions about the fundamental origins of BDs and how the mass, radius, and orbital properties of a BD reflects its formation and evolution.

\dbf{Mass, radius, age, and orbital properties are some of the key aspects that make up a complete understanding of the formation of transiting BDs. Traditionally, astronomers have defined BDs based on their ability to fuse deuterium and their inability to fuse hydrogen. Implicit in this definition is the assumption that BD formation is solely a function of mass. While it may be the case that mass is one of the more important factors in determining whether or not an object is a giant planet, BD, or low-mass star, there is a wealth of evolutionary information to be found in other basic properties.} 

\dbf{As we have explored here, the radius, age, and orbital eccentricity give us greater leverage towards understanding transiting BDs. We may combine orbital eccentricity and age with our knowledge of tidal timescales to examine the orbital history of a transiting BD. When the radius and age are used with the mass, we acquire a foothold into the mass-radius diagram for transiting BDs, where we may directly test the accuracy of substellar evolutionary models that seek to explain the underlying physics behind transiting BD formation. In this section, we will look at the population of transiting BDs and discuss how TOI-569b and TOI-1406b fit in to this picture.}

\begin{figure}[!ht]
\centering
\includegraphics[width=0.45\textwidth, trim={0.0cm 0.0cm 0.0cm 0.0cm}]{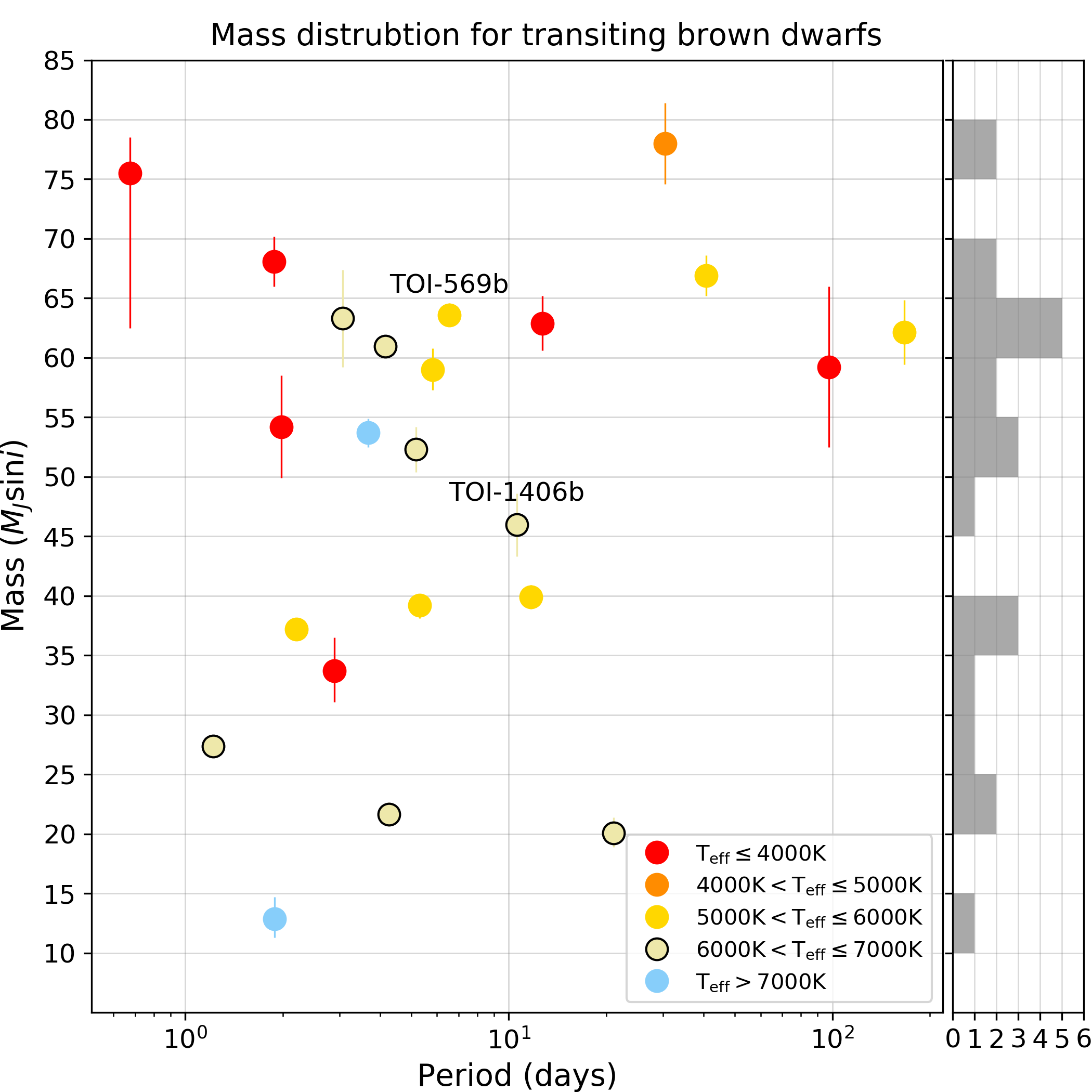}
\caption{Mass-period distribution of transiting BDs. The colors indicate the effective temperature of the host star of each BD.}\label{fig:massperiod}
\end{figure}

\subsection{Transiting brown dwarf host star distribution}
The mass distribution for the current population of transiting BDs is shown in Figure~\ref{fig:massperiod} with the effective temperature of the host star indicated by the colors of the points. From all published studies of transiting BDs to date, there is no obvious preference for a particular type of star to host transiting BDs. \dsout{This could be compared to the distribution of eclipsing primary stars in short-period binary systems and to hot Jupiter host stars in future works once the total number of transiting BDs increases.} Interestingly, we see that 6 transiting BDs (roughly 20\% of the transiting BD population, see Table~\ref{tab:bdlist}) are hosted by an M dwarf star. This is in contrast to hot Jupiters where only a few percent of the hot Jupiter population are found transiting M dwarf stars (e.g. Kepler-45b \cite{kepler45}, HATS-6b \cite{hats6}, WASP-80b \cite{wasp80}, NGTS-1b \cite{ngts1}). \dsout{It may be the case as TESS completes its primary mission that we will continue to discover more BDs around M dwarfs. If so, we will be challenged to explain why this is the case for BDs and less so for giant planets.} \dbf{By placing the transiting BD population in the context of eclipsing low-mass stars and hot Jupiters, we may also explore the idea that the scarcity of transiting BDs stems from them spanning the space between the tail ends of the distributions for companions that form like giant planets versus companions that form like low-mass stars. However, more transiting BD discoveries are needed for such studies to yield meaningful results.}

\subsection{Substellar isochrones and the mass-radius diagram}
\dbf{Here we will discuss how we use transiting BDs with well determined masses, radii, and ages to test the substellar isochrones from \cite{baraffe03} (for irradiated BDs) and \cite{saumon08} (for non-irradiated BDs). The \cite{baraffe03} models use the same input physics that \cite{chabrier97} used for main sequence stars. These are scaled appropriately in \cite{baraffe03} for low-mass stars and substellar objects down to $\lessapprox 1\mj$.  The way we test these models is by having independent measurements of a transiting BD's age, which comes from the age of it's host star (assuming the BD is the same age as its host star). We prioritize the use of stellar ages obtained through studies of clusters, asteroseismology, and gyrochronology, but with \textit{Gaia} DR2, we are able to reliably determine stellar properties to derive accurate ages with stellar isochrone models such as MIST. This increases the number of transiting BDs for which we have reliable and independently determined ages for comparison to substellar isochrones. This is important because we only know the companion as well as we know the host star and with \textit{Gaia} DR2, we can now know the host star to greater precision than ever before.}

\begin{figure}[!ht]
\centering
\includegraphics[width=0.45\textwidth, trim={0.0cm 0.0cm 0.0cm 0.0cm}]{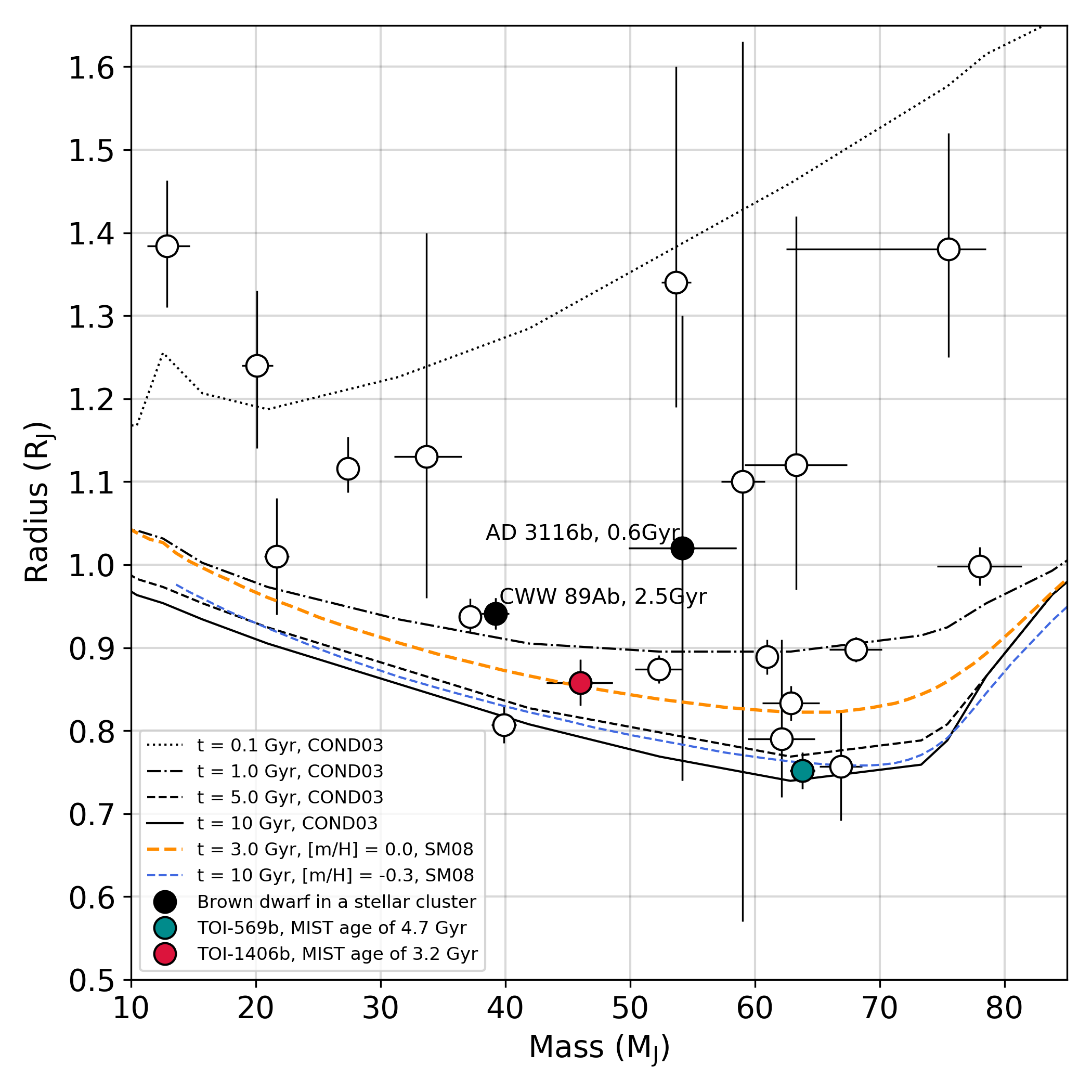}
\caption{Mass-radius diagram of transiting BDs featuring the COND03 and SM08 models. TOI-569b and TOI-1406b are shown as a cyan point and a red point, respectively. \dsout{Both models overestimate the age of TOI-569b yet seem to approximately estimate the age of TOI-1406b.} \dbf{Only 3 BDs that transit main sequence stars have ages constrained by stellar clusters or associations (AD 3116b in Preasepe, CWW 89Ab in Ruprecht 147, and RIK 72b in Upper Scorpius). Note RIK 72b is 5-10 Myr old \citep{david19_bd} and is not shown because its radius is $3.1\rj$. Also not shown are the eclipsing BDs in the BD binary system, 2M0535-05, located in the Orion Nebula Cluster with an age of 1-2Myr. Though TOI-569b and TOI-1406b are not in star clusters, we still have relatively precise ages for both from stellar isochrones of their host stars, and in a location on the mass-radius diagram where an age range of 5-10 Gyr results in little change in the models.}} \label{fig:massrad}
\end{figure}

The mass-radius diagram for transiting BDs is shown in Figure~\ref{fig:massrad}. All of the BDs on this diagram are necessarily transiting because it is through the transit method that we can measure the radius. However, even though a transit provides some measure of the radius, the measurement is not always very precise \dbf{(i.e. as precise as the measurement of the stellar radius). This is important because the radius of a BD changes drastically with its age \citep[see][]{baraffe03}}.\dsout{Exactly how precise can be quantified with the aid of substellar evolutionary models} \dbf{Substellar isochrones are challenging to test at ages beyond a few Gyr because the rate at which BD radii contract significantly decelerates, resulting in transiting BD radii approaching an asymptotic limit for the oldest systems (see the differences between the 3-10 Gyr substellar isochrones in Figure \ref{fig:massrad}).}

\dbf{Taking a more critical look at Figure \ref{fig:massrad}, we notice some interesting features. Of the most noticeable are the large ($> 10\%$) uncertainties on the radii of no fewer than 6 transiting BDs (AD 3116b, CoRoT-15b, CoRoT-33b, NGTS-7b, NLTT 41135b, and TOI-503b). These are the least informative data points in the substellar mass-radius diagram, especially for objects younger than 1 Gyr as the radius of a transiting BD changes rapidly at ages less than 1 Gyr. Notice also how the substellar isochrone models are mostly horizontal from roughly 20 to 70$\mj$. This means that testing the age of the isochrones is less sensitive to the precision on the mass of a transiting BD than it is to the precision on the radius.} 

\dbf{Interestingly, the oldest substellar isochrones are traced fairly well by a handful of transiting BDs. This suggests that the oldest substellar isochrones accurately predict the radii of transiting BDs that approach this asymptotic limit. Future works to improve on the \cite{baraffe03} (COND03) and \cite{saumon08} (SM08) models must consider the effects of metallicity for transiting BDs as this may be key to more finely distinguishing the older substellar isochrones from each other, especially in the asymptotic radius regime of 20 to 70$\mj$. The COND03 models do not explore a variety of different metallicity values as the SM08 models do, but the SM08 models do not consider the effects of irradiation like the COND03 models. Additionally, improvements may be made to BDs less massive than $20\mj$ as the input physical prescription from \cite{baraffe03} \& \cite{chabrier97} may not describe these BDs as well as they do more massive objects.}

\dsout{For example, we see from Figure~\ref{fig:massrad} that an uncertainty of $0.3\rj$ ($\sim$30\%) for any BD mass translates to an uncertainty of hundreds to thousands of Myr. These large (10-30\%) uncertainties in the radii typically arise from grazing transits of the BDs, as is the case with CoRoT-33b , NLTT 41135b , and TOI-503b. So, what may seem to be a reasonably precise radius uncertainty of 10\% actually turns out to be a rather large uncertainty in the age.} 

\subsubsection{TOI-569b and TOI-1406b in the mass-radius diagram}
\dbf{TOI-569b and TOI-1406b provide us the opportunity to test substellar isochrones older than 2.5 Gyr for the first time because we have accurate masses, radii, and ages traceable to stellar isochrones for their host stars. In this sense, we are using well-tested stellar isochrones to examine relatively untested substellar isochrones.} \dsout{We are fortunate that the transits of TOI-569b and TOI-1406b have low impact parameters, which means we have a nicely constrained radius measurements ($<5\%$ uncertainty).} \dsout{This means that these two new transiting BDs have precisely predicted ages from substellar evolutionary models. The Baraffe et al. 2003 (COND03) models consider the effects of irradiation on the BD's radius over time, while another set of models by Saumon et al. 2008 (SM08) does not and instead focuses on the effects of metallicity.}

Both the COND03 and the SM08 models seem to \dbf{slightly} overestimate the age of TOI-569b to be $\sim$10 Gyr compared to the age of the host star modelled from the \dbf{MIST} isochrones ($4.70\pm 1.30$ Gyr). \dbf{However, we note that the lower probability bimodal solution for TOI-569b favors a system age of $7.50\pm 1.80$ Gyr, which is in better agreement with the COND03 and SM08 models. The $\log{g}$ of TOI-569 also favors an older system.} 

\dbf{Something else worth note is that for a fixed BD mass and age, the radius increases with increasing metallicity \citep{burrows2011} and yet, TOI-569b has one of the smallest radii of all known transiting BDs with $  \rm [Fe/H]= +0.29$ dex (assuming it matches the host star). When referencing \cite{burrows2011}, Figure 1, we expect a change in the BD's metallicity ([Fe/H]) from +0.0 dex to +0.5 dex to result as a roughly 0.05-0.1$\rj$ increase in the radius of the BD. There is also an increase of about 0.05$\rj$ when transitioning from clear to cloudy atmospheric models for the BD. This is roughly a factor of two larger than our uncertainties on the radius TOI-569b ($R_{\rm BD}=0.75\pm 0.02\rj$).} For TOI-1406, we find that both the COND03 and SM08 models \dbf{do fairly well at} predicting the age of the system ($3.20^{+2.20}_{-1.60}$ Gyr). 

\dsout{Evidence for this may be seen in how both models overestimate the ages of CWW 89Ab and AD 3116b , which are two BDs with directly measured ages  via their association with stellar clusters/streams. Another way to convey this is to say that the COND03 models may underestimate the rate at which irradiated, short-period transiting BDs contract over long timescales.}

\subsection{Summary}
\dbf{TOI-569b and TOI-1406b are two newly discovered BDs that transit their host stars in nearly circular orbits. TOI-569 appears to be a slightly evolved G dwarf star with strong photometric modulation interpreted as evolving star spots on the surface of the star. We use the TESS and WASP light curves to extract an estimate for the rotation period of this star to be 13 days and determine that the star's rotational axis is marginally misaligned with the orbital inclination of TOI-569b. In contrast, TOI-1406 is an F star on the main sequence and with no noticeable photometric modulation over the sectors the star was observed by TESS. By comparing the ages of each system to a range of plausible circularization timescales, we find that we are not able to convincingly determine the orbital history of TOI-569 and that we can at least rule out significant high-eccentricity orbital evolution followed by tidal circularization for TOI-1406.}

\dbf{We demonstrate here how stellar isochrones can be used to test substellar isochrones. This is done by leveraging \textit{Gaia} DR2 for precise stellar parameters, which translate into better estimates of masses, radii, and ages, of transiting BD.} \dsout{Though the COND03 and SM08 models may not accurately estimate the ages of these BDs, we do have reasonable estimates of the ages of the host stars. So, we may compare the Jackson et al. 2008 models for circularization and synchronization timescales to these ages.} 

Ultimately, we find TOI-569b and TOI-1406b to be special in that they contribute new measurements to the still sparsely populated mass-radius diagram for transiting BDs. \dbf{In addition to providing some of the first examples of a test of the COND03 and SM08 models against stellar isochrones,} these systems also offer themselves as new data to examine circularization models. \dbf{As we build the population of transiting BDs, we will refine the predictive power of substellar isochrones and potentially turn them into tools useful in estimating the ages of transiting BDs.}

\section{Acknowledgements}
Funding for the TESS mission is provided by NASA's Science Mission directorate. This paper includes data collected by the TESS mission, which are publicly available from the Mikulski Archive for Space Telescopes (MAST). Resources supporting this work were provided by the NASA High-End Computing (HEC) Program through the NASA Advanced Supercomputing (NAS) Division at Ames Research Center for the production of the SPOC data products.

This work has made use of data from the European Space Agency (ESA) mission Gaia (https://www.cosmos.esa.int/gaia), processed by the Gaia Data Processing and Analysis Consortium (DPAC, https://www.cosmos.esa.int/web/gaia/dpac/consortium). Funding for the DPAC has been provided by national institutions, in particular the institutions participating in the Gaia Multilateral Agreement.

TWC acknowledges the efforts of the members of the TESS Followup Program and the Science Processing Operations Center in making the TESS data readily accessible for the analysis in this work.

Funding for this work is provided by the National Science Foundation Graduate Research Fellowship Program Fellowship (GRFP). This work makes use of observations from the LCOGT network.

AJM acknowledges support from the Knut \& Alice Wallenberg Foundation (project grant 2014.0017) and the Walter Gyllenberg Foundation of the Royal Physiographical Society in Lund.

CMP and MF gratefully acknowledge the support of the Swedish National Space Agency (DNR 163/16).

AJ and RB acknowledge support by the Ministry for the Economy, Development, and Tourism's Programa Iniciativa Cient\'{i}fica Milenio through grant IC\,120009, awarded to the Millennium Institute of Astrophysics (MAS). AJ acknowledges additional support from FONDECYT project 1171208.

\dbf{All authors especially acknowledge the efforts of the referee and thank them for their thoughtful and constructive feedback on this work.}

\facilities{TESS, Las Cumbres Observatory Global Telescope (LCOGT), SuperWASP, SOAR (HRCam), Gaia, Max Planck:2.2m (FEROS), CTIO:1.5m (CHIRON), ATT (optical echelle spectrograph), Euler:1.2m (CORALIE), WISE (infrared), CTIO:2MASS (optical, infrared)} 

{\dbf \software{{\tt EXOFASTv2} \citep{eastman2019}, {\tt pyanetti} \citep{2019MNRAS.482.1017B}, {\tt ceres} \citep{ceres}, LCO {\tt BANZAI} \citep{Collins:2017}, {\tt AstroImageJ} \citep{Collins:2017}} }

\bibliographystyle{aasjournal}
\bibliography{citations}

\begin{deluxetable}{cccccccccc}

\tabletypesize{\footnotesize}
\tablewidth{0pt}

 \tablecaption{List of published transiting brown dwarfs as of September 2019. \label{tab:bdlist}}

 \tablehead{
 \colhead{Name} & \colhead{$P$ (days)} & \colhead{$\rm M_{BD}/M_J$} & \colhead{$\rm R_{BD}/R_J$}& \colhead{e} & \colhead{$\rm M_\star/\mst$} &\colhead{$\rm R_\star/\rst$}& \colhead{$\rm T_{eff} (K)$}&\colhead{[Fe/H]} &\colhead{Reference}}
 \startdata 
 TOI-569b & 6.556 & $64.1 \pm 1.9$ & $0.75 \pm 0.02$ & $< 0.0035$ & $1.21 \pm 0.05$ & $1.47 \pm 0.03$ & $5768 \pm 110$ & $+0.29 \pm 0.09$ & this work\\ 
 TOI-1406b & 10.574 & $46.0\pm 2.7$ & $0.86\pm 0.03$ & $< 0.039$ & $1.18 \pm 0.09$ & $1.35\pm 0.03$ & $6290\pm 100$ & $-0.08 \pm 0.09$ & this work\\
 HATS-70b & 1.888 & $12.9\pm 1.8$ & $1.38\pm 0.08$ & $<0.18$ & $1.78 \pm 0.12$ & $1.88\pm 0.07$ & $7930\pm 820$ & $+0.04 \pm 0.11$ & 1\\
 KELT-1b & 1.218 & $27.4 \pm 0.9$ & $1.12 \pm 0.04$ & $0.01 \pm 0.01$ & $1.34 \pm 0.06$ & $1.47 \pm 0.05$ & $6516 \pm 49$ & $+0.05 \pm 0.08$ & 2\\
 NLTT 41135b & 2.889 & $33.7 \pm 2.8$ &  $1.13 \pm 0.27$ & $<0.02$ & $0.19 \pm 0.03$ & $0.21 \pm 0.02$ & $3230 \pm 130$ & $-0.25 \pm 0.25$ & 3\\
 LHS 6343c & 12.713 & $62.9 \pm 2.3$ & $0.83 \pm 0.02$ & $0.056 \pm 0.032$ & $0.37\pm 0.01$ & $0.38\pm 0.01$ & - & $+0.02 \pm 0.19$ & 4\\
 LP 261-75b & 1.882 & $68.1 \pm 2.1$ & $0.90 \pm 0.02$ & $<0.007$ & $0.30 \pm 0.02$ & $0.31 \pm 0.01$ & $3100 \pm 50$ & - & 5\\
 WASP-30b & 4.157 & $61.0 \pm 0.9$ & $0.89 \pm 0.02$ & 0 (adopted) & $1.17 \pm 0.03$ & $1.30 \pm 0.02$ & $6201 \pm 97$ & $-0.08 \pm 0.10$ & 6\\
 WASP-128b & 2.209 & $37.2 \pm 0.9$ & $0.94 \pm 0.02$ & $<0.007$ & $1.16 \pm 0.04$ & $1.15 \pm 0.02$ & $5950 \pm 50$ & $+0.01 \pm 0.12$ & 7\\
 CoRoT-3b & 4.257 & $21.7 \pm 1.0$ & $1.01 \pm 0.07$ & 0 (adopted) & $1.37 \pm 0.09$ & $1.56 \pm 0.09$ & $6740 \pm 140$ & $-0.02 \pm 0.06$ & 8\\
 CoRoT-15b & 3.060 & $63.3 \pm 4.1$ & $1.12 \pm 0.30$ & 0 (adopted) & $1.32 \pm 0.12$ & $1.46 \pm 0.31$ & $6350 \pm 200$ & $+0.10 \pm 0.20$ & 9\\
 CoRoT-33b & 5.819 & $59.0 \pm 1.8$ & $1.10 \pm 0.53$ & $0.070 \pm 0.002$ & $0.86 \pm 0.04$ & $0.94 \pm 0.14$ & $5225 \pm 80$ & $+0.44 \pm 0.10$ & 10\\
 Kepler-39b & 21.087 & $20.1 \pm 1.3$ & $1.24 \pm 0.10$ & $0.112 \pm 0.057$ & $1.29 \pm 0.07$ & $1.40 \pm 0.10$ & $6350 \pm 100$ & $+0.10 \pm 0.14$ & 11\\
 KOI-189b & 30.360 & $78.0 \pm 3.4$ & $1.00 \pm 0.02$ & $0.275 \pm 0.004$ & $0.76 \pm 0.05$ & $0.73 \pm 0.02$ & $4952 \pm 40$ & $-0.07 \pm 0.12$ & 12\\
 KOI-205b & 11.720 & $39.9 \pm 1.0$ & $0.81 \pm 0.02$ & $<0.031$ & $0.92 \pm 0.03$ & $0.84 \pm 0.02$ & $5237 \pm 60$ & $+0.14 \pm 0.12$ & 13\\
 KOI-415b & 166.788 & $62.1 \pm 2.7$ & $0.79 \pm 0.12$ & $0.689 \pm 0.001$ & $0.94 \pm 0.06$ & $1.15 \pm 0.15$ & $5810 \pm 80$ & $-0.24 \pm 0.11$ & 14\\
 EPIC 201702477b & 40.737 & $66.9 \pm 1.7$ & $0.76 \pm 0.07$ & $0.228 \pm 0.003$ & $0.87 \pm 0.03$ & $0.90 \pm 0.06$ & $5517 \pm 70$ & $-0.16 \pm 0.05$ & 15\\
 EPIC 212036875b & 5.170 & $52.3 \pm 1.9$ & $0.87 \pm 0.02$ & $0.132 \pm 0.004$ & $1.29 \pm 0.07$ & $1.50 \pm 0.03$ & $6238 \pm 60$ & $+0.01 \pm 0.10$ &  18, 21 \\
 AD 3116b & 1.983 & $54.2 \pm 4.3$ & $1.02 \pm 0.28$ & $0.146 \pm 0.024$ & $0.28 \pm 0.02$ & $0.29 \pm 0.08$ & $3200 \pm 200$ & $+0.16 \pm 0.10$ & 17 \\
 CWW 89Ab & 5.293 & $39.2 \pm 1.1$ & $0.94 \pm 0.02$ & $0.189 \pm 0.002$ & $1.10 \pm 0.05$ & $1.03 \pm 0.02$ & $5755 \pm 49$ & $+0.20 \pm 0.09$ & 16, 18 \\
 RIK 72b & 97.760 & $59.2 \pm 6.8$ & $3.10 \pm 0.31$ & $0.146 \pm 0.012$ & $0.44 \pm 0.04$ & $0.96 \pm 0.10$ & $3349 \pm 142$ & - & 19\\
 TOI-503b & 3.677 & $53.7 \pm 1.2$ & $1.34^{+0.26}_{-0.15}$ & 0 (adopted) & $1.80 \pm 0.06$ & $1.70 \pm 0.05$ & $7650 \pm 160$ & $+0.61 \pm 0.07$ & 22 \\
 NGTS-7Ab & 0.676 & $75.5^{+3.0}_{-13.7}$ & $1.38^{+0.13}_{-0.14}$ & 0 (adopted) & $0.48\pm 0.13$ & $0.61\pm 0.06$ & $3359\pm 106$ & - & 23\\
 2M0535-05a$\rm ^g$ & 9.779 & $56.7 \pm 4.8$ & $6.50 \pm 0.33$ & $0.323 \pm 0.006$ & - & - & - & - & 20\\
 2M0535-05b$\rm ^f$ & 9.779 & $35.6 \pm 2.8$ & $5.00 \pm 0.25$ & $0.323 \pm 0.006$ & - & - & - & - & 20\\
 \enddata
 \tablecomments{References: 1 - \cite{hats70b}, 2 - \cite{kelt1b}, 3 - \cite{irwin10}, 4 -  \cite{johnson11_bd}, 5 - \cite{irwin18}, 6 - \cite{wasp30b}, 7 - \cite{wasp128b}, 8 - \cite{corot3b}, 9 - \cite{corot15b}, 10 - \cite{corot33b}, 11 - \cite{kepler39}, 12 - \cite{diaz14}, 13 - \cite{diaz13}, 14 - \cite{moutou13}, 15 - \cite{bayliss16}, 16 - \cite{nowak17}, 17 - \cite{ad3116}, 18 - \cite{carmichael19}, 19 - \cite{david19_bd}, 20 - \cite{2M0535}, 21 - \cite{persson2019}, 22 - \cite{subjak2019}, 23 - \cite{jackman2019}}

\end{deluxetable}

\begin{deluxetable*}{lcccc}
\tablecaption{MIST median values and 68\% confidence interval for TOI-1406, created using {\tt EXOFASTv2} commit number 65aa674. Here, $\mathcal{U}$[a,b] is the uniform prior bounded between $a$ and $b$, and $\mathcal{G}[a,b]$ is a Gaussian prior of mean $a$ and width $b$. \dbf{We show $v\sin{I_\star}$ (taken to be the FWHM measured from CHIRON) only for convenient reference; {\tt EXOFASTv2} does not model FWHM for spectral line broadening.}}
\tablehead{\colhead{~~~Parameter} & \colhead{Units} & \colhead{Priors} & \multicolumn{2}{c}{Values}}
\startdata
\multicolumn{2}{l}{Stellar Parameters:}&\smallskip\\
~~~~$M_*$\dotfill &Mass (\msun)\dotfill & - &$1.18^{+0.08}_{-0.09}$\\
~~~~$R_*$\dotfill &Radius (\rsun)\dotfill & - &$1.35^{+0.03}_{-0.03}$\\
~~~~$L_*$\dotfill &Luminosity (\lsun)\dotfill & - &$2.56\pm0.15$\\
~~~~$\rho_*$\dotfill &Density (cgs)\dotfill & - &$0.68^{+0.07}_{-0.07}$\\
~~~~$\log{g}$\dotfill &Surface gravity (cgs)\dotfill & - &$4.252^{+0.037}_{-0.041}$\\
~~~~$T_{\rm eff}$\dotfill &Effective Temperature (K)\dotfill & $\mathcal{G}[6347,186]$ &$6290\pm100$\\
~~~~$[{\rm Fe/H}]$\dotfill &Metallicity (dex)\dotfill & $\mathcal{G}[-0.05,0.11]$ &$-0.08\pm0.09$\\
~~~~$Age$\dotfill &Age (Gyr)\dotfill & - &$3.20^{+2.20}_{-1.60}$\\
~~~~$EEP$\dotfill &Equal Evolutionary Point \dotfill & - &$377^{+40}_{-36}$\\
~~~~$A_V$\dotfill &V-band extinction (mag)\dotfill & $\mathcal{U}[0,0.08804]$ &$0.043^{+0.030}_{-0.029}$\\
~~~~$\sigma_{SED}$\dotfill &SED photometry error scaling \dotfill & - &$3.04^{+1.2}_{-0.73}$\\
~~~~$\varpi$\dotfill &Parallax (mas)\dotfill & $\mathcal{G}[2.3855,0.0291]$ &$2.386\pm0.029$\\
~~~~$d$\dotfill &Distance (pc)\dotfill & - & $419.1^{+5.2}_{-5.0}$\\
~~~~$v\sin{I_\star}$\dotfill &Projected equatorial velocity ($\rm km\, s^{-1}$)\dotfill & Not modelled & $12.91 \pm 0.24$\\
\\\multicolumn{2}{l}{Brown Dwarf Parameters:}& \smallskip\\
~~~~$P$\dotfill &Period (days)\dotfill & - &$10.57398^{+0.00060}_{-0.00059}$\\
~~~~$M_P$\dotfill &Mass (\mj)\dotfill  & - &$46.0^{+2.6}_{-2.7}$\\
~~~~$R_P$\dotfill &Radius (\rj)\dotfill & - & $0.86\pm0.03$\\
~~~~$T_C$\dotfill &Time of conjunction (\bjdtdb)\dotfill & - &$2458414.6065^{+0.0018}_{-0.0019}$\\
~~~~$a$\dotfill &Semi-major axis (AU)\dotfill  & - &$0.1010^{+0.0022}_{-0.0026}$\\
~~~~$i$\dotfill &Orbital inclination (Degrees)\dotfill & - &$87.70^{+0.19}_{-0.20}$\\
~~~~$e$\dotfill &Eccentricity \dotfill  & - &$0.026^{+0.013}_{-0.010}$\\
~~~~$ecos{\omega_*}$\dotfill & \dotfill & - &$-0.0160^{+0.0079}_{-0.0071}$\\
~~~~$esin{\omega_*}$\dotfill & \dotfill & - &$0.017^{+0.017}_{-0.015}$\\
~~~~$T_{eq}$\dotfill &Equilibrium temperature (K)\dotfill  & - &$1108^{+18}_{-17}$\\
~~~~$K$\dotfill &RV semi-amplitude ($\rm m\, s^{-1}$)\dotfill  & - &$3720^{+120}_{-130}$\\
~~~~$logK$\dotfill &Log of RV semi-amplitude \dotfill & - &$3.570^{+0.014}_{-0.015}$\\
~~~~$R_P/R_*$\dotfill &Radius of planet in stellar radii \dotfill  & - &$0.0654\pm0.0011$\\
~~~~$a/R_*$\dotfill &Semi-major axis in stellar radii \dotfill  & - &$16.11^{+0.56}_{-0.58}$\\
~~~~$\delta$\dotfill &Transit depth (fraction)\dotfill  & - &$0.00428\pm0.00014$\\
~~~~$\tau$\dotfill &Ingress/egress transit duration (days)\dotfill  & - &$0.0180^{+0.0016}_{-0.0014}$\\
~~~~$b$\dotfill &Transit Impact parameter \dotfill  & - &$0.648^{+0.033}_{-0.036}$\\
~~~~$\rho_P$\dotfill &Density (cgs)\dotfill & -  &$90^{+11}_{-10.}$\\
~~~~$logg_P$\dotfill &Surface gravity \dotfill & -  &$5.190^{+0.040}_{-0.042}$\\
~~~~$M_P\sin i$\dotfill &Minimum mass (\mj)\dotfill & -  &$45.9^{+2.6}_{-2.7}$\\
~~~~$M_P/M_*$\dotfill &Mass ratio \dotfill & -  &$0.0372^{+0.0016}_{-0.0015}$\\
\hline
\\\multicolumn{2}{l}{Wavelength Parameters:}&TESS band\smallskip\\
~~~~$u_{1}$\dotfill &linear limb-darkening coeff \dotfill  &$0.224\pm0.050$\\
~~~~$u_{2}$\dotfill &quadratic limb-darkening coeff \dotfill  &$0.299\pm0.050$\\
\\\multicolumn{2}{l}{RV Parameters:}&ANU&CHIRON\smallskip\\
~~~~$\gamma_{\rm rel}$\dotfill &Relative RV Offset ($\rm m\, s^{-1}$)\dotfill &$-15490^{+200}_{-240}$&$-15461^{+70}_{-68}$\\
~~~~$\sigma_J$\dotfill &RV Jitter ($\rm m\, s^{-1}$)\dotfill &$380^{+400}_{-260}$&$182^{+85}_{-56}$\\
~~~~$\sigma_J^2$\dotfill &RV Jitter Variance \dotfill &$150000^{+460000}_{-130000}$&$33000^{+38000}_{-17000}$\\
\\\multicolumn{2}{l}{Transit Parameters:}&TESS \smallskip\\
~~~~$\sigma^{2}$\dotfill &Added Variance \dotfill &$-0.000000154^{+0.000000011}_{-0.000000010}$\\
~~~~$F_0$\dotfill &Baseline flux \dotfill &$0.999996\pm0.000013$\\
\enddata
\end{deluxetable*}\label{tab:exofast_toi1406}

\begin{deluxetable*}{lcccccc}
\tablecaption{\dbf{MIST} median values and 68\% confidence interval for TOI-569, created using {\tt EXOFASTv2} commit number 65aa674. The most likely values (probability of 0.73) and the ones we report for this system are shown in boldface. Here, $\mathcal{U}$[a,b] is the uniform prior bounded between $a$ and $b$, and $\mathcal{G}[a,b]$ is a Gaussian prior of mean $a$ and width $b$. \dbf{We show $v\sin{I_\star}$ (measured from CHIRON) only for convenient reference; {\tt EXOFASTv2} does not model $v\sin{I_\star}$.}}
\tablehead{\colhead{~~~Parameter} & \colhead{Units} & \colhead{Priors} & \colhead{\bf Most likely values} & \multicolumn{2}{c}{Less likely values}}
\startdata
\multicolumn{2}{l}{Stellar Parameters:}& & $\rm \bf Prob.=0.73$& $\rm Prob.=0.27$\smallskip\\
~~~~$M_*$\dotfill &Mass (\msun)\dotfill & - &$\bf 1.21\pm 0.05$ & $1.10^{+0.03}_{-0.05}$\\
~~~~$R_*$\dotfill &Radius (\rsun)\dotfill & - &$\bf 1.48\pm0.03$ & $1.48\pm0.03$\\
~~~~$L_*$\dotfill &Luminosity (\lsun)\dotfill  & - &$\bf 2.15^{+0.15}_{-0.12}$ & $2.09\pm 0.1$\\
~~~~$\rho_*$\dotfill &Density (cgs)\dotfill  & - &$\bf 0.54\pm 0.04$ &$0.48\pm 0.03$\\
~~~~$\log{g}$\dotfill &Surface gravity (cgs)\dotfill  & - & $\bf 4.19\pm 0.03$&$4.14\pm 0.02$\\
~~~~$T_{\rm eff}$\dotfill &Effective Temperature (K)\dotfill  & $\mathcal{G}[5699,106]$ &$\bf 5768^{+110}_{-92}$ & $5720^{+94}_{-85}$\\
~~~~$[{\rm Fe/H}]$\dotfill &Metallicity (dex)\dotfill  & $\mathcal{G}[+0.23,0.10]$ &$\bf +0.29^{+0.09}_{-0.08}$ &$+0.23\pm 0.09$\\
~~~~$Age$\dotfill &Age (Gyr)\dotfill  & - &$\bf 4.70\pm 1.3$& $7.50^{+1.80}_{-1.20}$\\
~~~~$A_V$\dotfill &V-band extinction (mag)\dotfill  & $\mathcal{U}[0,1.1749]$ &$\bf 0.067^{+0.083}_{-0.049}$&$0.052^{+0.074}_{-0.038}$\\
~~~~$\sigma_{SED}$\dotfill &SED photometry error scaling \dotfill  & - &$\bf 2.99^{+1.10}_{-0.71}$&$3.08^{+1.40}_{-0.78}$\\
~~~~$\varpi$\dotfill &Parallax (mas)\dotfill  & $\mathcal{G}[6.3723,0.0306]$ &$\bf 6.374\pm 0.031$&$6.374^{+0.031}_{-0.030}$\\
~~~~$d$\dotfill &Distance (pc)\dotfill  & - &$\bf 156.88\pm0.75$&$156.83\pm0.75$\\
~~~~$v\sin{I_\star}$\dotfill &Projected equatorial velocity ($\rm km\, s^{-1}$)\dotfill & Not modelled & $\bf 5.30 \pm 0.50$ & $5.30 \pm 0.50$\\
\\\multicolumn{2}{l}{Brown Dwarf Parameters:}& &\smallskip\\
~~~~$P$\dotfill &Period (days)\dotfill & - &$\bf 6.55604^{+0.00016}_{-0.00015}$&$6.55603^{+0.00016}_{-0.00015}$\\
~~~~$M_P$\dotfill &Mass (\mj)\dotfill  & - &$\bf 64.1^{+1.9}_{-1.4}$ & $59.6^{+1.1}_{-1.7}$\\
~~~~$R_P$\dotfill &Radius (\rj)\dotfill & - &$\bf 0.75\pm0.02$ & $0.76\pm 0.02$\\
~~~~$T_C$\dotfill &Time of conjunction (\bjdtdb)\dotfill & - &$\bf 2458523.09192^{+0.00070}_{-0.00069}$ & $2458523.09199\pm 0.00070$\\
~~~~$a$\dotfill &Semi-major axis (AU)\dotfill  & - &$\bf 0.07428\pm 0.00059$ & $0.07207^{+0.00047}_{-0.00069}$\\
~~~~$i$\dotfill &Orbital inclination (Degrees)\dotfill & - &$\bf 85.37^{+0.13}_{-0.11}$ & $85.15^{+0.13}_{-0.12}$\\
~~~~$e$\dotfill &Eccentricity \dotfill  & - &$\bf 0.0017^{+0.0018}_{-0.0012}$ & $0.0017^{+0.0018}_{-0.0012}$\\
~~~~$ecos{\omega_*}$\dotfill & \dotfill & - & $\bf 0.0002^{+0.0015}_{-0.0011}$ & $0.0002^{+0.0015}_{-0.0011}$\\
~~~~$esin{\omega_*}$\dotfill & \dotfill & - & $\bf 0.0005^{+0.0022}_{-0.0012}$ & $0.0005^{+0.0022}_{-0.0012}$\\
~~~~$T_{eq}$\dotfill &Equilibrium temperature (K)\dotfill  & - &$\bf 1227^{+13}_{-12}$ & $1237^{+12}_{-13}$\\
~~~~$K$\dotfill &RV semi-amplitude ($\rm m\, s^{-1}$)\dotfill  & - &$\bf 5884\pm17$ & $5884^{+18}_{-17}$\\
~~~~$logK$\dotfill &Log of RV semi-amplitude \dotfill & - &$\bf 3.7697\pm0.0013$ &$3.7697\pm0.0013$\\
~~~~$R_P/R_*$\dotfill &Radius of planet in stellar radii \dotfill & -  &$\bf 0.05217^{+0.00094}_{-0.00091}$&$0.05258^{+0.0010}_{-0.00096}$\\
~~~~$a/R_*$\dotfill &Semi-major axis in stellar radii \dotfill  & - &$\bf 10.81^{+0.22}_{-0.21}$ &$10.44^{+0.21}_{-0.20}$\\
~~~~$\delta$\dotfill &Transit depth (fraction)\dotfill  & - &$\bf 0.002721^{+0.00010}_{-0.000096}$&$0.002765^{+0.00011}_{-0.000100}$\\
~~~~$\tau$\dotfill &Ingress/egress transit duration (days)\dotfill  & - &$\bf 0.0214^{+0.0015}_{-0.0013}$ &$0.0230^{+0.0014}_{-0.0013}$\\
~~~~$b$\dotfill &Transit Impact parameter \dotfill  & - &$\bf 0.8739^{+0.0082}_{-0.0084}$&$0.8820^{+0.0068}_{-0.0075}$\\
~~~~$\rho_P$\dotfill &Density (cgs)\dotfill  & - &$\bf 187^{+17}_{-16}$ &$169^{+16}_{-15}$\\
~~~~$logg_P$\dotfill &Surface gravity \dotfill  & - &$\bf 5.444^{+0.029}_{-0.031}$ &$5.412^{+0.027}_{-0.026}$\\
~~~~$M_P\sin i$\dotfill &Minimum mass (\mj)\dotfill  & - &$\bf 63.6\pm 1.0$&$59.9^{+0.8}_{-0.9}$\\
~~~~$M_P/M_*$\dotfill &Mass ratio \dotfill  & - &$\bf 0.05043^{+0.00097}_{-0.00050}$&$0.05195^{+0.00045}_{-0.00039}$\\
\hline
\\\multicolumn{2}{l}{Wavelength Parameters:}&I-band&TESS band\smallskip\\
~~~~$u_{1}$\dotfill &linear limb-darkening coeff \dotfill &$0.292\pm0.050$&$0.335\pm0.049$\\
~~~~$u_{2}$\dotfill &quadratic limb-darkening coeff \dotfill &$0.260\pm0.050$&$0.269\pm0.049$\\
\\\multicolumn{2}{l}{RV Parameters:}&CHIRON&CORALIE&FEROS\smallskip\\
~~~~$\gamma_{\rm rel}$\dotfill &Relative RV Offset ($\rm m\, s^{-1}$)\dotfill &$71964^{+14}_{-16}$&$73413^{+17}_{-18}$&$73402^{+63}_{-55}$\\
~~~~$\sigma_J$\dotfill &RV Jitter ($\rm m\, s^{-1}$)\dotfill &$37^{+23}_{-17}$&$37^{+27}_{-16}$&$70^{+210}_{-70}$\\
~~~~$\sigma_J^2$\dotfill &RV Jitter Variance \dotfill &$1420^{+2300}_{-1000}$&$1390^{+2800}_{-920}$&$5000^{+76000}_{-7100}$\\
\\\multicolumn{2}{l}{Transit Parameters:}&LCOGT UT 2019-04-15 (I-band)&TESS \\
~~~~$\sigma^{2}$\dotfill &Added Variance \dotfill &$0.00000202^{+0.00000030}_{-0.00000026}$&$0.0000001792^{+0.0000000063}_{-0.0000000062}$\\
~~~~$F_0$\dotfill &Baseline flux \dotfill &$0.99981\pm0.00013$&$1.0000143\pm0.0000053$\\
\enddata
\end{deluxetable*}\label{tab:exofast_toi569}

\end{document}